\documentclass[aps,twocolumn,pre,superscriptaddress, floatfix]{revtex4-1}

\usepackage{xcolor,hyperref}
\hypersetup{
   colorlinks,
   linkcolor={blue!50!black},
   citecolor={blue!50!black},
   urlcolor={blue!80!black}
}

\usepackage{amsmath}
\usepackage{amssymb}
\usepackage{color}
\definecolor{mypink2}{RGB}{219, 48, 122}

\usepackage{graphicx}

\DeclareMathAlphabet{\mathitbf}{OML}{cmm}{b}{it}

\newcommand{\zv}{\mathitbf z}

\newcommand{\xv}{\mathitbf x}
\newcommand{\uv}{\mathitbf u}
\newcommand{\dv}{\mathitbf d}

\newcommand{\piv}{\mathBold\pi}
\newcommand{\psiv}{\mathBold\psi}

\newcommand{\calBold}[1]{\mbox{\boldmath${\cal #1}$}}
\newcommand{\mathBold}[1]{\mbox{\boldmath$#1$}}
\newcommand{\dbar}{{\,\mathchar'26\mkern-12mu d}}

\newcommand{\sFrac}[2]{{\textstyle\frac{#1}{#2}}}

\usepackage{chemformula}

\begin{document}

\title{Mechanical disorder of  sticky-sphere glasses. I. Effect of attractive interactions}

\author{Karina Gonz\'alez-L\'opez}
\affiliation{Institute for Theoretical Physics, University of Amsterdam, Science Park 904, Amsterdam, Netherlands}
\author{Mahajan Shivam}
\affiliation{School of Physical and Mathematical Sciences, Nanyang Technological University, Singapore 637371, Singapore}
\author{Yuanjian Zheng}
\affiliation{School of Physical and Mathematical Sciences, Nanyang Technological University, Singapore 637371, Singapore}
\author{Massimo Pica Ciamarra}
\affiliation{School of Physical and Mathematical Sciences, Nanyang Technological University, Singapore 637371, Singapore}
\affiliation{CNR-SPIN, Dipartimento di Scienze Fisiche, Università di Napoli Federico II, I-80126 Napoli, Italy}
\author{Edan Lerner}
\email{e.lerner@uva.nl}
\affiliation{Institute for Theoretical Physics, University of Amsterdam, Science Park 904, Amsterdam, Netherlands}

\begin{abstract}
Recent literature indicates that attractive interactions between particles of a dense liquid play a secondary role in determining its bulk mechanical properties. Here we show that, in contrast with their apparent unimportance to the bulk mechanics of dense liquids, attractive interactions can have a major effect on macro- and microscopic elastic properties of glassy solids. We study several broadly-applicable dimensionless measures of stability and mechanical disorder in simple computer glasses, in which the relative strength of attractive interactions --- referred to as `glass stickiness' --- can be readily tuned. We show that increasing glass stickiness can result in the decrease of various quantifiers of mechanical disorder, on both macro- and microscopic scales, with a pair of intriguing exceptions to this rule. Interestingly, in some cases strong attractions can lead to a reduction of the number density of soft, quasilocalized modes, by up to an order of magnitude, and to a substantial decrease in their core size, similar to the effects of thermal annealing on elasticity observed in recent works. Contrary to the behavior of canonical glass models, we provide compelling evidence indicating that the stabilization mechanism in our sticky-sphere glasses stems predominantly from the self-organized depletion of interactions featuring large, \emph{negative} stiffnesses. Finally, we establish a fundamental link between macroscopic and microscopic quantifiers of mechanical disorder, which we motivate via scaling arguments. Future research directions are discussed. 
\end{abstract}

\maketitle

\section{introduction}
\label{sec:intro}
\vspace{-0.3cm}
One of the key challenges in glass physics is establishing robust relations between structure and dynamics~\cite{falk_langer_stz,gotze2008complex,sgr_prl_1997,rob_jack_PRE_2007,harrowell_isoconfiguration,widmer2008irreversible,manning2011,Eran_Jim_STZ_PRE3_2009,smarajit_review,paddy_huge_review_2015,cge_paper,Tanaka_PRX_2018,Liesbeth_mct_review_2018,Barrat_elastoplastic_review_2018,david_huge_collaboration,Kurchan_Levine_2010,Tanaka_PRX_2018,Tanaka_Tong_prl_2020}. The various approaches to assessing glassy matter's structure can be roughly partitioned into two schools: approaches that focus on positional disorder~\cite{gotze2008complex,paddy_huge_review_2015,free_volume_1961,Tanaka_Tong_prl_2020}, and those that focus on \emph{mechanical} disorder \cite{falk_langer_stz,Barrat_elastoplastic_review_2018,widmer2008irreversible,manning2011,Eran_Jim_STZ_PRE3_2009}. Mechanical disorder refers to the various forms of mechanical fluctuations that are seen to emerge in structural glasses, ranging from localized soft spots~\cite{Schober_Laird_numerics_PRL,widmer2008irreversible,manning_defects,modes_prl_2016} to mesoscopic fluctuations in elastic moduli fields~\cite{Schirmacher_prl_2007,mizuno_mossa_barrat_pre_2013,everyone,mossa_prb_2020}.

The future utility of glasses in technological innovations and industrial applications depends both upon improving glass formation processes \cite{LOFFLER2003529,Park2005,schroers_review,Schroers_bmgs,Li2016}, 
and upon understanding the relations between glasses' composition --- which, in turn, determines the nature of interactions between the constituent particles --- and their mechanical properties \cite{composition1,composition2,composition3,wang_review_2012,Launey4986}. While the process of glass formation and glass-forming ability have been the subject of enormous research efforts \cite{johnson_gfa_apl_2010,Zhang_gfa_2015}, the effects of different types of microscopic interactions on the emergent mechanical disorder of glasses have not been fully explored yet.

One field in which substantial success has been achieved in understanding relations between microscopic observables in disordered materials, and their macro- and microscopic elasticity, is that of unjamming \cite{ohern2003,liu_review,van_hecke_review,everyone}. The unjamming scenario precisely describes the changes in microscopic elasticity --- in terms of e.g.~characteristic frequency- and length-scales featured by vibrational modes \cite{matthieu_PRE_2005,Silbert_prl_2005,mw_EM_epl,eric_boson_peak_emt,new_variational_argument_epl_2016} --- and the changes in macroscopic elasticity --- in terms of e.g.~various elastic moduli ratios \cite{Ellenbroek_2009,everyone,stefanz_pre_2019} --- upon reducing the degree of connectedness of the network of~interactions between the constituents of an amorphous solid. 

While the phenomenology associated with the unjamming scenario has been largely understood, less focus has been dedicated by the materials-physics community to studying the effects of the \emph{form} of interactions between the constituent particles of a glass on its elasticity, in systems largely removed from the unjamming point \cite{Braier1990,Sciortino2002,Tanguy_pre_2010_vary_lambda_sw,potential_itamar_pre_2011,itamar_brittle_to_ductile_pre_2011,wang_review_2012,beltukov_sw_pre_2016,massimo_soft_matter_2013,shi_intrinsic_ductility,experimental_inannealability,smarajit_ductile_brittle_soft_matter_2016,Tanguy_pre_2017_silicate_glass}. In Ref.~\cite{boring_paper} an effort was made to identify, trace and quantitatively compare between some of the trends observed in macroscopic elasticity of simple computer glasses, under changes of various parameters of interest. Here we adopt a similar strategy, except that we focus in particular on the role of attractive terms in pairwise interaction potentials on the elastic properties of the resulting glasses, in system for which the unjamming scenario is irrelevant~\cite{footnote}. To this aim, we employ two computer-glass models put forward in Refs.~\cite{potential_itamar_pre_2011,mie_potential}, in which particles interact via a pairwise potential whose attractive term can be straightforwardly tuned. We study these models to shed light on the question: how do attractive interactions affect elastic properties, mechanical disorder and stability of glassy solids?

Comparing the elastic properties of different types of glasses, with the aim of cleanly distilling the effects emanating from different features in the form of interparticle interactions, is generally not straightforward to accomplish. The difficulty to do so stems from the history-dependence that glasses' properties are notorious for, that might hinder comparison of elastic properties between different types of glasses on an equal footing.  In computer experiments, this difficulty can be circumvented; several works have shown that elastic properties of glasses instantaneously quenched from parent temperatures $T_{\mbox{\footnotesize p}}$ --- larger than some onset temperature $T_{\mbox{\footnotesize on}}$ --- become nearly independent of $T_{\mbox{\footnotesize p}}$ \cite{Sastry1998, SASTRY2002267, Ashwin2004, cge_paper, LB_modes_2019, boring_paper}. The onset temperature $T_{\mbox{\footnotesize on}}$ is found to be roughly twice the computer glass transition temperature $T_g$, defined here as the temperature at which the primary relaxation time is of order $10^4\tau_0$, where $\tau_0$ is a microscopic timescale. Here we follow Refs.~\cite{boring_paper, modes_prl_2020} and exploit this feature of glass-forming models, in order to achieve a meaningful, quantitative comparison between them. We create glassy samples by quenching high-temperature liquid states equilibrated at temperatures that are roughly a factor of 4 higher than their respective computer $T_g$, and therefore much larger than the onset temperature $T_{\mbox{\footnotesize on}}$. The effects of varying $T_{\mbox{\footnotesize p}}$ from high-temperature liquid states to supercooled viscous liquid states, on micro- and macroscopic elasticity of computer glasses featuring attractive interactions, is the focus of an accompanying paper \cite{inannealability}. 

In addition to studying the effects of strong attractions between a glass constituents on its macroscopic elastic properties, here we focus much of our attention to the investigation of the effects of attractive interactions on what is referred to here as \emph{microscopic} elasticity: the statistical, structural and energetic properties of soft, quasilocalized modes \cite{Schober_Laird_numerics_PRL,modes_prl_2016,modes_prl_2018,modes_prl_2020} (QLMs). These nonphononic soft excitations are observed in essentially all glasses quenched from a melt \cite{modes_prl_2016,modes_prl_2018,modes_prl_2020}, and are microscopic in nature: they feature a disordered core of linear size $\xi_g$ \cite{modes_prl_2016,pinching_pnas} (defined and studied in detail in what follows), which is typically on the order of a few interparticle distances, and is decorated by algebraically-decaying Eshelby-like far fields \cite{modes_prl_2016}.

The prevalence of QLMs or lack thereof have been shown to be extremely sensitive to the degree of supercooling of parent configurations from which glassy samples are instantaneously quenched \cite{SchoberOligschleger1996,cge_paper,LB_modes_2019,pinching_pnas,corrado_dipole_statistics_2020}; in Ref.~\cite{pinching_pnas} it was shown that in glassy samples that underlie deeply supercooled equilibrium configurations, the number density ${\cal N}(T_{\mbox{\footnotesize p}})$ of soft QLMs can decrease by two orders of magnitude compared to that found in glasses quenched from high parent temperatures $T_{\mbox{\footnotesize p}}$. In the same work it was also shown that the core size of QLMs decreases, and their typical frequency increases, with deeper supercooling of glasses' ancestral equilibrium states. 

For these aforementioned reasons, the statistical and structural properties of QLMs can be associated with the notion of \emph{mechanical disorder}, and are therefore studied carefully throughout this work under variations of the relative strength of attractive interactions between the constituent particles of our computer glasses. Our study echoes some of the findings of Ref.~\cite{itamar_brittle_to_ductile_pre_2011}, in which the same glass-forming model as (one of the two models) employed here was investigated, but a different set of observables were considered; in what follows we explain in detail the similarities and differences between our work and Ref.~\cite{itamar_brittle_to_ductile_pre_2011}. 

Here we establish that the relative strength of attractive pairwise interactions --- defined and referred to below as `glass stickiness' --- can be a major contributor to glass stability. We show that increasing glass stickiness by employing steeper attractive terms in the interaction potential, or, alternatively, by decompressing a glass, can lead to a decrease in QLMs' core size, accompanied by the stiffening of QLMs' characteristic frequency scale. In some particularly interesting cases --- discussed in detail in what follows ---, we find that the number density of QLMs decreases by up to an order of magnitude compared to that found in more generic glass models \cite{modes_prl_2020}. These effects, as well as the effects of stickiness on macroscopic moduli, are reminiscent in essence, as well as in magnitude, to the trends seen in thermally-annealed \cite{cge_paper,LB_modes_2019,pinching_pnas} or otherwise-stabilized \cite{fsp} glasses. 

\begin{figure}[ht!]
\centering
  \includegraphics[width=1.0\linewidth]{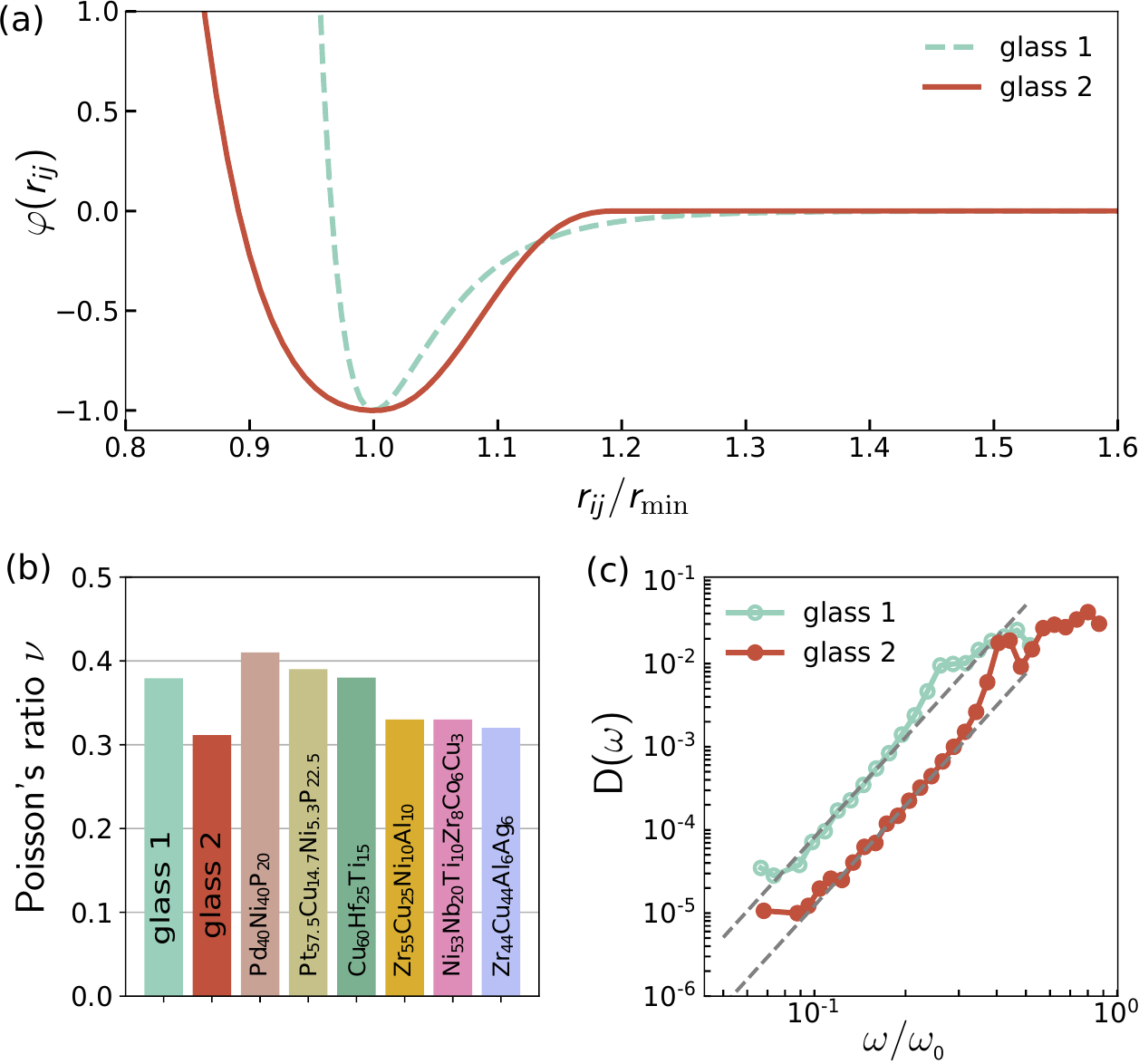}
\label{fig:fig1}
\end{figure}

Another key conclusion of our study is that small changes in the shape of the attractive part of the employed pairwise potentials can lead to dramatic changes in the emergent micro- and macroscopic elastic properties of their resulting glasses. Consider, for example, the two pairwise potentials shown in Fig.~\ref{fig:fig1}(a), referred to as `glass 1' and `glass 2'. These two potentials do not differ significantly from each other in terms of the form of their respective attractive parts. Interestingly, despite this similarity, their resulting glasses' Poisson's ratio decreases by $\approx$ 18\% between `glass 1' to `glass 2' (see Fig.~\ref{fig:fig1}b), and the density per frequency $D(\omega)$ of low-frequency, quasilocalized modes of these models differs substantially, by a factor of $\approx$ 6.5 (see Fig.~\ref{fig:fig1}(c)). In what follows we provide evidence indicating that these differences stem from the self-organized abundance or dearth of interactions featuring destabilizing \emph{negative} stiffnesses.

This work is structured as follows; in Sec.~\ref{sect:models} we reintroduce the two model systems employed in our investigation, and explain how the three sets of ensembles of computer glasses analyzed below were created. Sect.~\ref{sec:stickiness} defines the notion of glass stickiness, and shows how it behaves in our computer glasses. In Secs.~\ref{sect:macro} and~\ref{sect:meso} we report the results of our investigation of the glass-stickiness-dependence of macro- and microscopic elasticity, respectively. In Sec.~\ref{sect:discussion} we hold several discussions in which the results of our computer-experiments are interpreted, explained, and discussed in the context of other recent work, together with various additional analyses. In Sec.~\ref{sect:summary} we briefly summarize the main results of the manuscript, and Sec.~\ref{sect:outlook} provides an outlook of future research directions. Precise definitions of the observables considered in our work, some of their measurement methods, and additional supporting data pertaining to some of the discussions, are all deferred to the appendixes.

\vspace{-1.8cm}

\begin{figure}[h!]
\centering
  \includegraphics[width=0.95\linewidth]{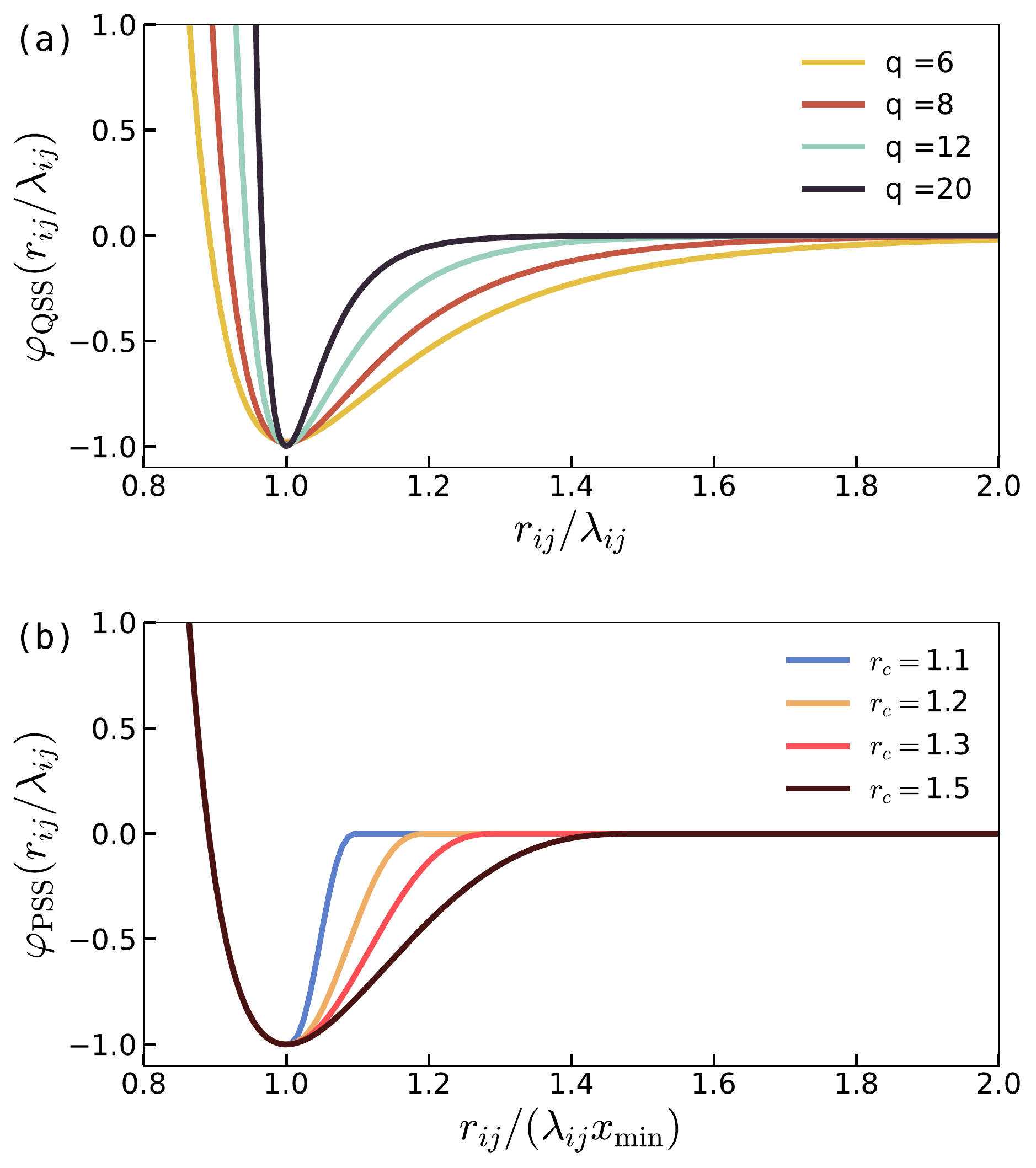}
\caption{\footnotesize (a) The pairwise potential $\varphi_{\mbox{\tiny QSS}}$ of the QSS model (see Eq.~(\ref{eq:qstickypot})), plotted for different powers $q$ as indicated by the legend. (b) The pairwise potential $\varphi_{\mbox{\tiny PSS}}$ of the PSS model (see Eq.~(\ref{eq:sspotential})), for the selected dimensionless cutoffs $r_c$ indicated by the legend, expressed in terms of the dimensionless location of the minimum of the potential $x_{\rm min}\!=\!2^{1/6}$. \label{fig:potentials}}
\end{figure}

\vspace{-1.5cm}

\section{Computer glass models, glass ensembles and units}
\label{sect:models}


In this study we employ two computer glass models of 50:50 binary mixtures of `small' and `large' particles of equal mass $m$ in three dimensions (3D) at fixed volume $V$, interacting via radially-symmetric, pairwise interaction potentials $\varphi_{ij}(r_{ij})$, such that the total potential energy is written as $U\!=\!\sum_{i<j}\varphi_{ij}(r_{ij})$. With these two models, we created \emph{three} sets of glass ensembles, as detailed in Sec.~\ref{sec:protocol}. 


\vspace{-0.13cm}

\subsection{Computer glass models}
\vspace{-0.2cm}

\subsubsection{q-sticky-spheres}

\vspace{-0.2cm}

The first model employed is a generalized Lennard-Jones potential, also known as Mie potential \cite{mie_potential}, in which the strength of the attractive and repulsive interactions can be tuned via an exponent $q$. The supercooled-liquid dynamics of this model was studied in Ref.~\cite{sri_potential}. The pairwise potential of this model reads
\vspace{-0.1cm}

\begin{eqnarray}\label{eq:qstickypot}
    \varphi_{\mbox{\tiny QSS}}(r_{ij}) = \!\varepsilon\! \bigg[ \big(\sFrac{\lambda_{ij}}{r_{ij}}\big)^{2q}\!\! - 2\big(\sFrac{\lambda_{ij}}{r_{ij}}\big)^{q} + \sum\limits_{\ell=0}^{3} c_{\mbox{\tiny $2\ell$}} \big(\sFrac{r_{ij}}{\lambda_{ij}}\big)^{2\ell} \bigg]
\end{eqnarray}
for $r_{ij}/\lambda_{ij}\!<\!x_c$, and $\varphi_{\mbox{\tiny QSS}}(r_{ij})\!=\!0$ for $r_{ij}/\lambda_{ij}\!\ge\!x_c$. Here $\varepsilon$ is a microscopic energy scale, $\lambda_{ij}$ is a length parameter (see below), and $c_{\mbox{\tiny $2\ell$}}(q,x_c)$ are the $q$- and $x_c$-dependent coefficients (reported in Appendix.~\ref{app:coefficients}) that ensure the first and second derivatives of $\varphi_{\mbox{\tiny QSS}}$ with respect to the pairwise distance $r_{ij}$ are continuous at the dimensionless cutoff $x_c$. We chose the parameters $q\!=\!6,8,12,20$ and $x_c\!=\!3.0,2.5,2.0,1.8$, respectively. The length parameters are expressed in terms of the `small-small' interaction length $\lambda_{\mbox{\tiny small}}^{\mbox{\tiny small}}$, with $\lambda_{\mbox{\tiny small}}^{\mbox{\tiny large}}\!=\!1.18\lambda_{\mbox{\tiny small}}^{\mbox{\tiny small}}$ and $\lambda_{\mbox{\tiny large}}^{\mbox{\tiny large}}\!=\!1.4\lambda_{\mbox{\tiny small}}^{\mbox{\tiny small}}$. The potential as given by Eq.~(\ref{eq:qstickypot}) is plotted in Fig.~\ref{fig:potentials}a. We refer to this model as \emph{q-sticky-spheres}.

\vspace{-0.5cm}

\subsubsection{Piecewise-sticky-spheres}
\vspace{-0.2cm}

The second model employed in this work is a Lennard-Jones-like potential, first introduced in Ref.~\cite{potential_itamar_pre_2011}, in which the repulsive part of the pairwise potential is identical to the canonical Lennard Jones (LJ) potential, but the range --- and therefore also the strength --- of the attractive part can be readily modified, as shown in Fig.~\ref{fig:potentials}b. The supercooled liquid dynamics of this model was studied in Ref.~\cite{Massimo_supercooled_PRL}. The piecewise pairwise potential of this model reads
\begin{widetext}

\begin{equation}
    \varphi_{\mbox{\tiny PSS}}(r_{ij}) \!=\!
\left\{
\begin{array}{cc}
\!\!4\varepsilon \bigg[ \big(\frac{\lambda_{ij}}{r_{ij}}\big)^{12} - \big(\frac{\lambda_{ij}}{r_{ij}}\big) ^{6} \bigg],
     &  \frac{r_{ij}}{\lambda_{ij}} <  x_{\mbox{\tiny min}}  \\
\varepsilon \bigg[a\big(\frac{\lambda_{ij}}{r_{ij}}\big)^{12} -b\big(\frac{\lambda_{ij}}{r_{ij}}\big)^{6} + \sum\limits_{\ell=0} ^{3}  c_{\mbox{\tiny $2\ell$}} \big(\frac{r_{ij}}{\lambda_{ij}}\big)^{2\ell} \bigg] , & x_{\mbox{\tiny min}}\!\le\! \frac{r_{ij}}{\lambda_{ij}}< x_c\\
0\,,  & x_c \le \frac{r_{ij}}{\lambda_{ij}}
\end{array}
\right. ,
 \label{eq:sspotential}
\end{equation}
\end{widetext}
where $\varepsilon$ is a microscopic energy scale, $x_{\mbox{\tiny min}},x_c$ are the (dimensionless) locations of the minimum of the LJ potential and modified cutoff, respectively, and the $\lambda_{ij}$'s are the same length parameters as detailed for the QSS model above. In what follows, we express the dimensionless cutoff $x_c$ in terms of $x_{\mbox{\tiny min}}\!=\!2^{1/6}$, for simplicity, by defining $r_c\!\equiv\!x_c/x_{\mbox{\tiny min}}$; $r_c$ serves as one of the key control parameters in our investigation, as explained below. The coefficients $a,b,\{c_{\mbox{\tiny $2\ell$}}\}$ are chosen such that the attractive and repulsive parts of $\varphi_{\mbox{\tiny PSS}}$, and its first two derivatives, are continuous at $x_{\mbox{\tiny min}}$ and at $x_c$, see  Table~\ref{tab:CSSoeff} in Appendix~\ref{app:coefficients} for the coefficients' numerical values. We refer to this model as \emph{piecewise-sticky-spheres}.

\begin{figure*}
\centering
  \includegraphics[width=0.895\linewidth]{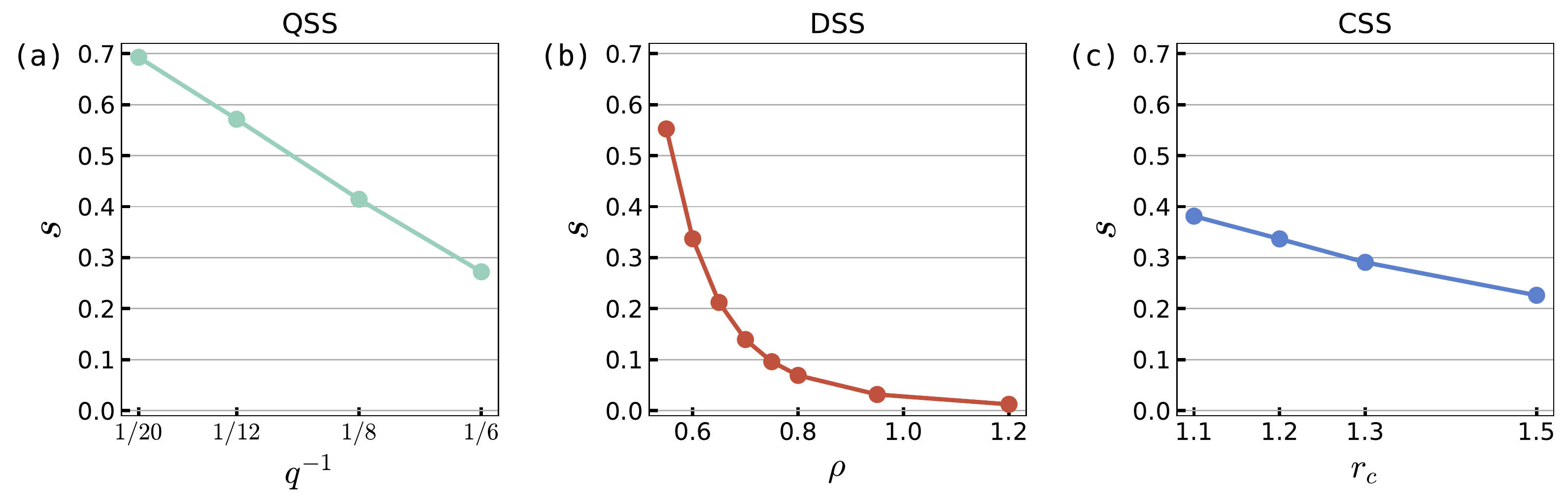}
\caption{\footnotesize `Glass stickiness' $s$ --- defined in Eq.~\eqref{eq:stickiness} --- \emph{vs}.~the respective control parameters of the (a) QSS, (b) DSS, and (c) CSS glass ensembles. In what follows, we maintain the same order between figure panels and different glass ensembles as presented here.}
\label{fig:stickiness}
\end{figure*}

\subsection{Computer glass ensembles and their preparation protocol}\label{sec:protocol}

We created 3 sets of ensembles of glasses, as follows:
\begin{enumerate}
    \item Q-Sticky-Spheres (employing $\varphi_{\mbox{\tiny QSS}}$; see Eq.~(\ref{eq:qstickypot})), with $q=6,8,12,20$ and $x_c\!=\!3.0,2.5,2.0,1.8$, respectively; varying $q$ in this model at fixed density leads to the variation of the dimensionless pressure $p/p_0$ where $p_0$ is a characteristic pressure scale defined precisely in Appendix.~\ref{app:p_0}. For this reason, we tuned the density $\rho\!\equiv\!mN/V$ such that $p/p_0\!\approx\!0.05$ for all $q$, resulting in $\rho\!=\!0.80, 0.74, 0.698, 0.677$ for $q\!=\!6,8,12,20$, respectively. These ensembles are referred to as QSS. 
    \item Piecewise-Sticky-Spheres (i.e., employing $\varphi_{\mbox{\tiny PSS}}$; see Eq.~(\ref{eq:sspotential})), varying Density (referred to as DSS): piecewise-sticky-spheres glasses with fixed cutoff $r_c\!=\!1.2$ and varying density $\rho\!=\!0.55,$  0.60, 0.65, 0.70, 0.75, 0.80, 0.95, 1.20. 
    \item Piecewise-Sticky-Spheres (i.e.~employing $\varphi_{\mbox{\tiny PSS}}$; see Eq.~(\ref{eq:sspotential})), varying Cutoff (referred to as CSS): piecewise-sticky-spheres glasses with fixed density $\rho\!=\!0.60$ and varying cutoff $r_c\!=\!1.1$, 1.15, 1.2, 1.3, 1.5. To avoid data cluttering, we present data only for $r_c\!=\!1.15$ in Figs.~\ref{fig:prefactor},\ref{fig:all_core_length_observables},\ref{fig:wg_N_QLE},\ref{fig:ag_vs_chig},\ref{fig:ag_vs_nu}. 
\end{enumerate}

\begin{table}[!ht]
\caption{\label{sys_sizes}
\footnotesize System and ensemble sizes of glassy samples generated for this work. }
\begin{ruledtabular}
\begin{tabular}{cccccc}
ensemble & $N$ & $n$ \\ 
\hline
QSS $(\forall q)$ & 10,648 & 1,300 \\
\hline
DSS $(\rho=0.55)$ & 3,000 & 9,200 \\
\hline
DSS $(\rho \in [0.60,0.95])$ & 10,000 & 1,000\\
\hline
DSS $(\rho=1.2)$ & 16,000 & 1,000\\
\hline
CSS $(r_{c}=1.1)$ & 3,000 & 50,200 \\
\hline
CSS $(r_{c}=1.15)$ & 3,000  & 40,000 \\
\hline
CSS $(r_{c} \in [1.2,1.3])$ & 3,000 & 9,200\\
\hline
CSS $(r_{c}=1.5)$ & 10,000 & 3,000\\

\end{tabular}
\end{ruledtabular}
\end{table}

Ensembles of independent glassy samples were created as follows; we first equilibrated each system in the high temperature liquid state using standard molecular dynamics (MD) simulations. The equilibration temperature for each system was chosen to be at least 4 times higher than $G_\infty a_0^3/30$, where $a_0\!\equiv\!(V/N)^{1/3}$ is the typical interparticle distance, and $G_{\infty}$ is the high-temperature limit of the sample-to-sample mean of the shear modulus of underlying inherent states (glasses) found by a quick quench. We empirically observe that $G_\infty a_0^3/30$ is a rough estimation of the computer glass transition temperature in many numerical models of supercooled liquids, which motivates our choice. The temperature was kept constant by employing the Berendsen thermostat \cite{Berendsen}. After equilibration, the energy is minimized instantaneously using a standard conjugate gradient algorithm. We followed this protocol to generated $n$ independent glasses for each ensemble. The total number of samples generated for each model, and the system sizes employed, are all detailed in Table~\ref{sys_sizes}. The dimensionless pressures of the QSS and CSS glass ensembles are shown in Fig.~\ref{fig:pressures} in Appendix~\ref{app:p_0}. We finally note that the minimal coordination number over all of our glass ensembles, pertaining to the CSS ensemble with $r_c\!=\!1.1$, is $\approx\!10.5$, i.e.~very far from the Maxwell threshold of 6 (in 3D). Consequently, none of the effects we observe in this work are related to the unjamming transition~\cite{ohern2003,liu_review,van_hecke_review,everyone}.

\subsection{Units}
In what follows, we report all lengths in terms of the characteristic interparticle distance $a_0\!\equiv\!(V/N)^{1/3}$, and all frequencies in terms of $\omega_0\!\equiv\!c_s/a_0$, where the (ensemble dependent) speed of shear waves is given by $c_s\!=\!\sqrt{G/\rho}$ with $G$ denoting the (ensemble dependent) shear modulus and $\rho\!\equiv\!mN/V$ denoting the mass density.

\section{glass stickiness}
\label{sec:stickiness}

As stated in the introduction, we consider here computer glasses in which the strength of attractive interactions can be tuned. In order to assess the effective relative strength of attractive vs.~repulsive forces in a meaningful way, we define a glass's `stickiness' as follows; for each particle $i$ we identify the largest repulsive pairwise \emph{force} $f^{\mbox{\tiny rep}}_i\!=\!\mbox{max}_j (-\varphi'_{ij})$ and largest (most negative) attractive pairwise force $f^{\mbox{\tiny att}}_i\!=\!\mbox{max}_j(\varphi'_{ij})$. The stickiness $s$ of a glass is then defined as
\begin{equation}\label{eq:stickiness}
    s \equiv \frac{\mbox{mean}_i f^{\mbox{\tiny att}}_i}{\mbox{mean}_i f^{\mbox{\tiny rep}}_i}\,,
\end{equation}
where mean$_i$ denotes a mean over particles $i$. Clearly $s\!=\!0$ for purely-repulsive systems, and $s\!>\!0$ once attractive interactions are present. 

Importantly, we note that stickiness is an \emph{emergent} property of a glass, affected by its self-organized structure, which is controlled, in turn, by features of the pairwise interactions, and by mechanical-equilibrium and spatial-confinement constraints to which the glass is subjected. In Fig.~\ref{fig:stickiness} we show our measurements of glass stickiness for our three sets of glass ensembles. It is clear that glass stickiness can be manipulated either by varying features of the employed pairwise potentials (cf.~QSS and CSS models), or by compressing or decompressing a system with a fixed pairwise potential, as shown for the DSS system in Fig.~\ref{fig:stickiness}b. This latter route to controlling glass stickiness might be relevant experimentally \cite{XIAO2015190}.

\begin{figure*}[ht!]
\includegraphics[width=0.895\linewidth]{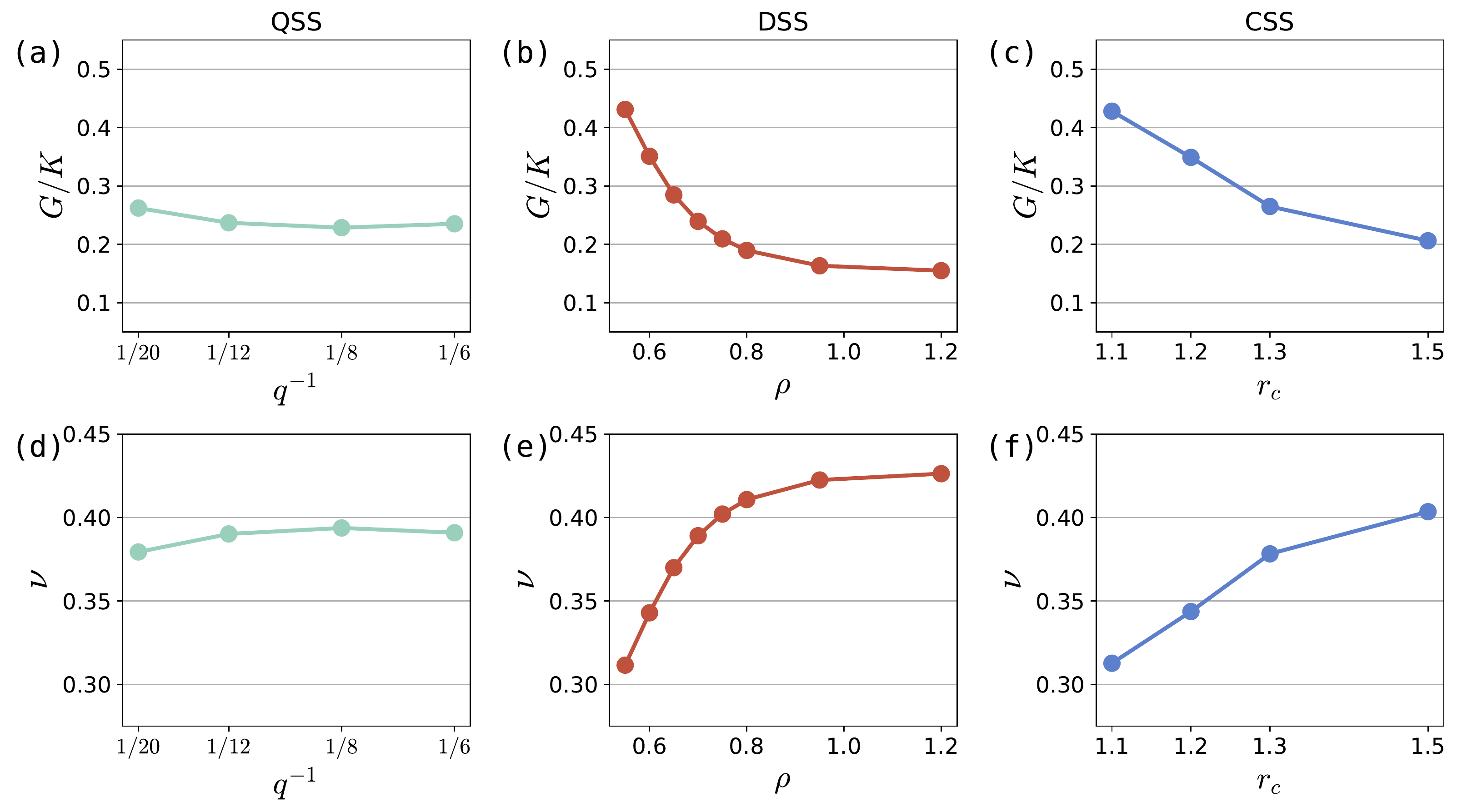}
\caption{\footnotesize Panels (a)-(c) show the sample-to-sample mean athermal shear to bulk modulus ratio $G/K$, and panels (d)-(f) show the Poisson's ratio $\nu\!\equiv\!(3\!-\!2G/K)/(6\!+\!2G/K)$, plotted as a function of the key control parameter pertaining to each glass ensemble. 
\label{fig:g_over_k_and_nu}}
\end{figure*}


\begin{figure*}[ht!]
\includegraphics[width=0.87\linewidth]{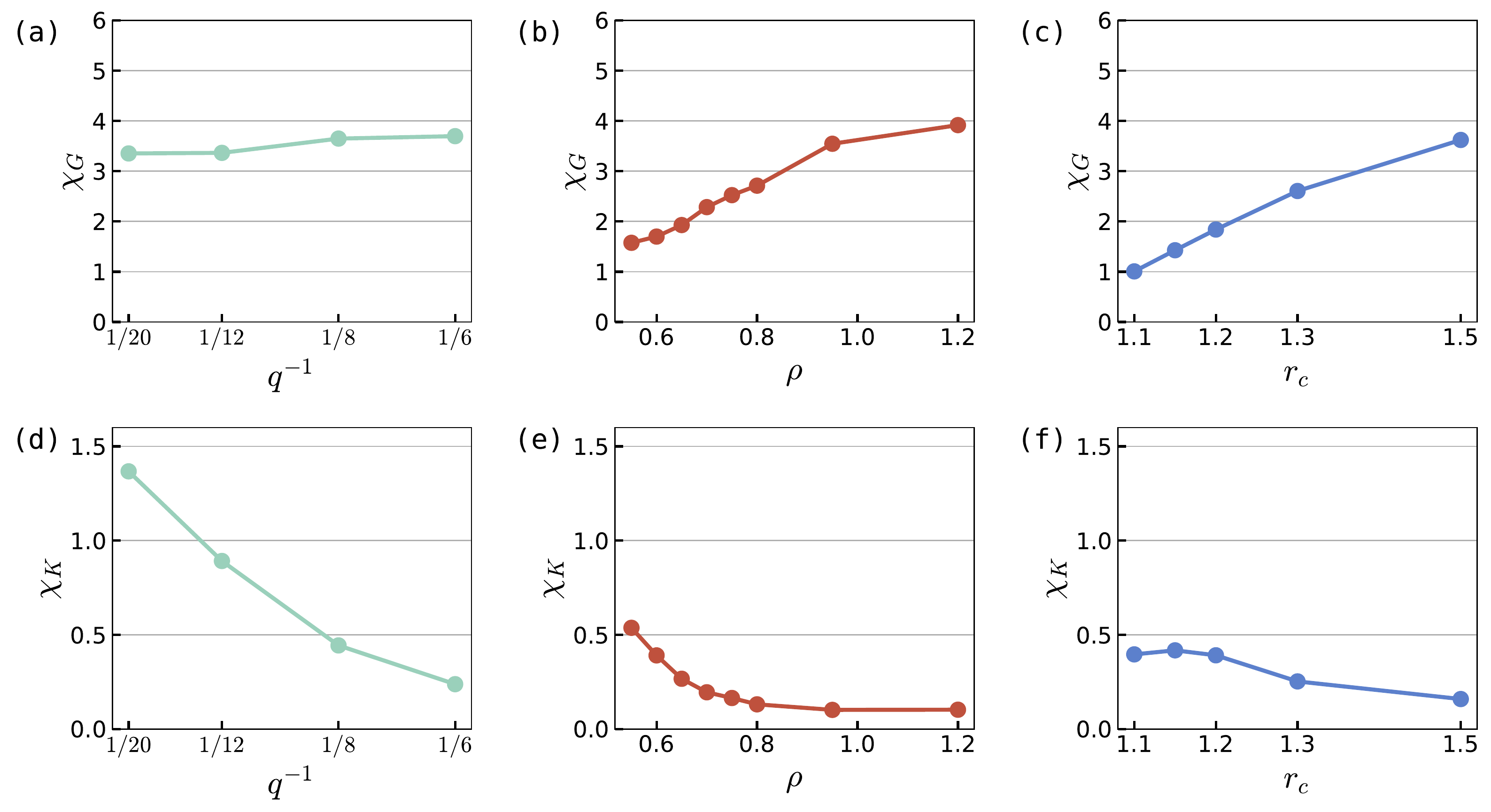}
\caption{\footnotesize Relative sample-to-sample fluctuations of shear and bulk moduli as captured by $\chi_{_G}$ (a)-(c), and $\chi_{_K}$ (d)-(f), respectively, defined in Eq.~(\ref{eq:chi}); see text and Appendix~\ref{app:chi} for more details. \label{fig:chi}}
\end{figure*}

\section{Effect of glass stickiness on macroscopic elasticity}
\label{sect:macro}

\subsection{Elastic moduli and Poisson's ratio}

We start our investigation of macroelasticity by studying the sample-to-sample averages of athermal shear to bulk moduli ratio $G/K$, and Poisson's ratio $\nu$ (see Appendix~\ref{appx:macro} for precise definitions), of the QSS, DSS, and CSS ensembles described in the previous section, as a function of their respective control parameters. The results are shown in Fig.~\ref{fig:g_over_k_and_nu}; we observe that $G/K$ increases significantly in the DSS and CSS ensembles by decreasing $\rho$ and $r_{c}$, respectively, with a total variation of $\approx\!180\%$ in the former case. In addition, the high density plateau $G/K\!\approx\!0.15$ observed for the DSS glasses is in excellent agreement with the values measured for computer glasses featuring purely repulsive, inverse-power-law pairwise potentials \cite{boring_paper}. This indicates that at high densities the relative strength of attractive interactions is too small to play an important role in determining the elastic properties of the DSS glasses, as expected.

\begin{figure*}[!ht]
\centering
  \includegraphics[width=0.87\linewidth]{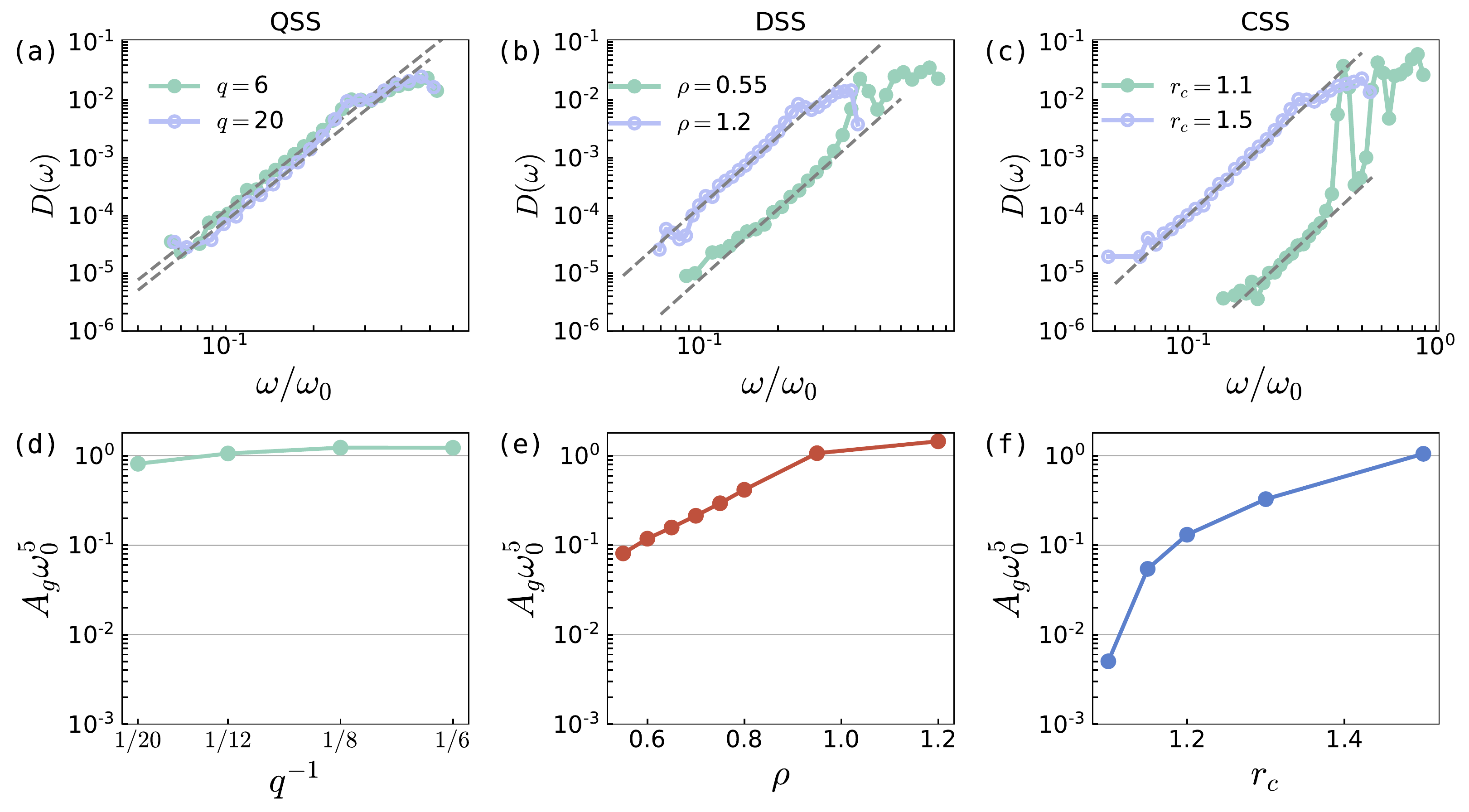}
\caption{\footnotesize Panels (a), (b), (c) show the vDOS for the highest and lowest values of the respective key control parameter, in the QSS, DSS, and CSS systems. The dashed lines represent our fits of the low-frequency power law tails, according to Eq.~\eqref{d_of_omega}. In panels (d)-(f) we report the fitted prefactors $A_{g}$ for all systems \emph{vs.}~their respective control parameter, see text for further discussion.
\label{fig:prefactor}}
\end{figure*}

In contrast, we observe in Figs.~\ref{fig:g_over_k_and_nu}a and \ref{fig:g_over_k_and_nu}d that $G/K$, and therefore also $\nu$, are largely insensitive to varying glass stickiness in the QSS model, indicating that glass stickiness alone is not predictive of elastic properties. In particular, we show in Fig.~\ref{fig:fig1} that the QSS pairwise interaction with $q\!=\!20$ (referred to as `glass 1' in that figure) has a very comparable attractive term to that of the CSS interaction with $r_c\!=\!1.2$ (referred to as `glass 2' in Fig.~\ref{fig:fig1}), but nevertheless features a much \emph{smaller} $G/K$ ratio, and therefore also a larger Poisson's ratio $\nu$. The observed difference between the large-glass-stickiness elastic properties of the QSS ensemble, and those of the DSS and CSS ensembles, will be discussed further in what follows. 

\vspace{-0.57cm}
\subsection{Elastic moduli's relative fluctuations}

In addition to ensemble-averaged values of the shear and bulk moduli, and of their ratio, we consider next two dimensionless, $N$-independent measures of their sample-to-sample fluctuations, defined respectively as 
\begin{equation}\label{eq:chi}
    \chi_{_G}\! =\! \frac{\sqrt{N\langle (G - \langle G \rangle)^2 \rangle}}{\langle G \rangle} \ \  \mbox{and} \ \  \chi_{_K} \!=\! \frac{\sqrt{N\langle (K - \langle K \rangle)^2 \rangle}}{\langle K \rangle}\,,
\end{equation} 
where $\langle\bullet\rangle$ denotes an ensemble average. These observables can be considered as quantifiers of mechanical disorder \cite{fsp,boring_paper}; similar quantifiers based on spatial (coarse-grained) statistics rather than sample-to-sample statistics were put forward by Schirmacher \cite{Schirmacher_2006,Schirmacher_prl_2007,Marruzzo2013} in the context of the vibrational density of states and transport properties of glasses, and also discussed in Ref.~\cite{riggleman_soft_matter_2010,mizuno_mossa_barrat_pre_2013,barrat_pnas_2014,mossa_prb_2020}. Recently, $\chi_{_G}$ was shown in Ref.~\cite{scattering_prl_2020} to predict long-wavelength wave attenuation rates in computer glasses. Appendix~\ref{appx:macro} provides details about how $\chi_{_G}$ and $\chi_{_K}$ were estimated for our glass ensembles. 

Our results are shown in Fig.~\ref{fig:chi}; we find stark differences between the way the two quantifiers $\chi_{_G}$ and $\chi_{_K}$ depend on each model's key control parameter. Remarkably, $\chi_{_G}$ is roughly independent of $q$ in the QSS system, in line with the underwhelming variation of $G/K$ with $q$ as seen in Fig.~\ref{fig:g_over_k_and_nu}. At the same time $\chi_{_G}$ significantly decreases upon increasing glass stickiness in the DSS and CSS ensembles, suggesting the increased mechanical stability of those glasses \cite{boring_paper,fsp}.

In contrast, $\chi_{_K}$ appears to depend strongly on $q$ in the QSS glasses -- it \emph{grows} by roughly a factor of 6 with increasing $q$. Similarly, increasing glass stickiness in the DSS and CSS ensembles leads to an \emph{increase} in the sample-to-sample fluctuations of $K$ as captured by $\chi_{_K}$, opposite to the \emph{decrease} we observe in $\chi_{_G}$ upon increasing glass stickiness. We note, importantly, that the value of $\chi_{_K}$ remains substantially smaller than $\chi_{_G}$ under all considered scenarios. These trends and their origins are further discussed in Sec.~\ref{sect:discussion}.


\section{Effect of glass stickiness on microscopic elasticity}
\label{sect:meso}

As discussed in the Introduction, the macroscopic mechanical stability of disordered solids is often related to the statistical and energetic properties of low-frequency, nonphononic, quasilocalized vibrational modes that emerge from the microscopic disorder and glassy structural frustration~\cite{cge_paper,fsp,LB_modes_2019,Ozawa6656,pinching_pnas,corrado_dipole_statistics_2020}. In this Section we thus focus on the microscopic elasticity of the glasses generated by the three protocols described in Sec.~\ref{sec:protocol}. Precise definitions of the observables studied can be found in Appendix~\ref{app:micro}.

\begin{figure*}
\includegraphics[width=0.9\linewidth]{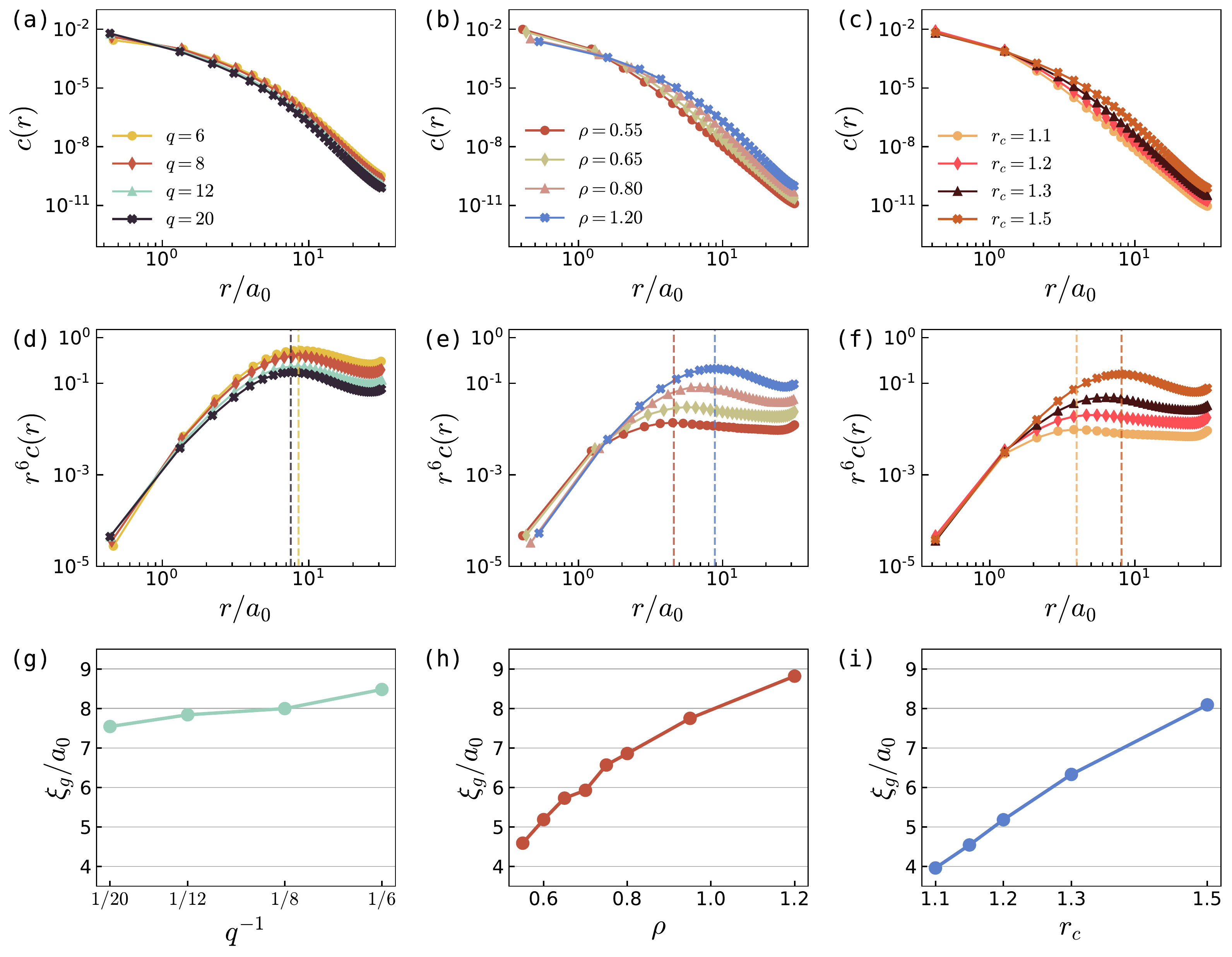}
\caption{\footnotesize Panels (a)-(c) show the response functions $c(r)$ to local force perturbation (see precise definition in Appendix~\ref{xi_g_and_omega_g_appendix}), calculated in glasses of $\approx$ 250K particles.  Panels (d)-(f) show the products $r^{6}c(r)$, that features a crossover to the expected continuum scaling $c(r)\!\sim\!r^{-6}$ \cite{breakdown} beyond a length $\xi_g$, defined here as the maximum of the products $r^{6}c(r)$. $\xi_g$ is extracted as explained in Appendix~\ref{xi_g_and_omega_g_appendix}, and reported in panels (g)-(i).
\label{fig:all_core_length_observables}}
\end{figure*}

\subsection{Density per frequency of quasilocalized modes}

Under a certain set of circumstances \cite{modes_prl_2016,modes_prl_2018,modes_prl_2020,phonon_widths,pinching_pnas}, the vibrational density of states (vDOS) of structural glasses has been shown to grow from zero frequency as
\begin{equation}\label{d_of_omega}
   D(\omega)\!=\!A_{g}\omega^{4}\,,
\end{equation}
independent of spatial dimension \cite{modes_prl_2018,atsushi_high_dimension_dos}, glass history \cite{pinching_pnas}, or interaction details \cite{modes_prl_2020}. The vibrational modes that populate this asymptotic scaling regime were shown to be \emph{quasilocalized}; they feature a disordered core of size $\xi_g$, decorated by algebraically-decaying fields $\sim\!r^{1-\dbar}$ \cite{modes_prl_2016,atsushi_core_size_pre}. The prefactor $A_{g}$, featuring units of frequency$^{-5}$, has been argued \cite{cge_paper,pinching_pnas} to encompass information regarding both the characteristic frequency $\omega_g$ of quasilocalized modes (QLMs), and their number density ${\cal N}$ (per particle), discussed further below. 

Here we measure the prefactor $A_{g}$ by fitting the scaling law Eq.~(\ref{d_of_omega}) to the low-frequency tail of the vDOS; see Figs.~\ref{fig:prefactor}(a)-\ref{fig:prefactor}(c) for some examples. Figures \ref{fig:prefactor}(d)-\ref{fig:prefactor}(f) of Fig.~\ref{fig:prefactor} report the extracted prefactors $A_g$ for our different glass ensembles, as a function of their respective control parameters. In the QSS system, we observe that $A_{g}$ remains almost constant, independent of the exponent $q$, consistent with and similar to the near independence of $G/K$ on $q$. In contrast with this behavior, $A_{g}$ varies by an order of magnitude in the DSS glasses, and by almost two orders of magnitude in the CSS glasses, upon increasing glass stickiness. We find the lowest prefactor to be featured by the CSS system with $r_{c}\!=\!1.1$, for which $A_g\approx\!4\!\times\!10^{-3}$.

\begin{figure*}[ht!]
\centering
  \includegraphics[width=0.9\linewidth]{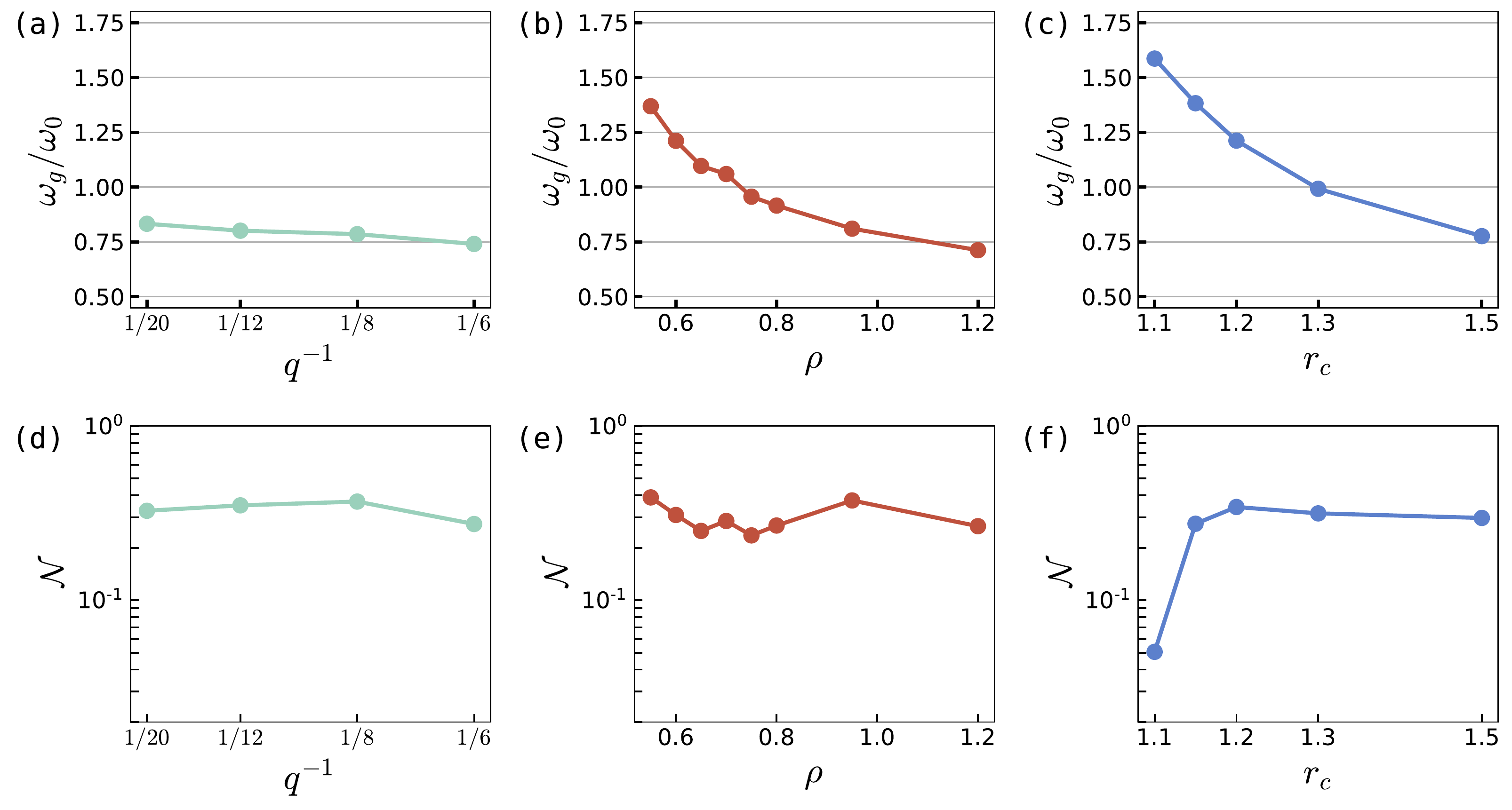}
\caption{\footnotesize (a)-(c) The characteristic frequency $\omega_g$ of QLMs, obtained via Eq.~(\ref{eq:omega_g}), see text for discussion. Panels (d)-(f) show the resulting number density of QLMs ${\cal{N}}\!=\!A_g\omega_g^{5}$, estimated for our glass ensembles as a function of their respective key control parameters.
\label{fig:wg_N_QLE}}
\end{figure*}

Similar analyses were carried out in Refs.~\cite{itamar_brittle_to_ductile_pre_2011,smarajit_ductile_brittle_soft_matter_2016}, where the same CSS model was studied. In Ref.~\cite{itamar_brittle_to_ductile_pre_2011} it was argued that the prefactor $B$ (in the notations of that work) of a (power-law) distribution of `plastic modes'' (presumably the same objects referred to here as QLMs) stiffnesses was inferred by subtracting the Debye contribution \cite{kittel2005introduction} from the eigenvalue distribution of the Hessian matrix. The result of this inference was that $B$ varies over a factor of $\approx$ 4.5, by varying $r_c$ from 1.3 to 2.4, in good agreement with our measurements, (cf.~Fig.~\ref{fig:prefactor}), and notice that we probed smaller $r_c$'s. Different from our analysis here, in the analysis of Ref.~\cite{itamar_brittle_to_ductile_pre_2011} the exponent of the plastic modes' stiffness distribution --- known now to be universal \cite{modes_prl_2016,modes_prl_2018,modes_prl_2020} (however see Refs.~\cite{protocol_prerc,finite_size_effect_modes_2020}) --- was treated as a fitting parameter (whose values are not reported), and the explicit finite-size-effects seen in the Debye contribution \cite{phonon_widths} were not accounted for. Nevertheless, we stress that the trend we observe for $A_g$ with varying glass stickiness has been pointed out first in Ref.~\cite{itamar_brittle_to_ductile_pre_2011}.

We next turn to examining the effect of glass stickiness on the core size of QLMs.


\subsection{Core-size of quasilocalized modes}

Following Ref.~\cite{pinching_pnas}, the length $\xi_{g}$ that represents the (linear) core size of QLMs can be obtained by considering the characteristic spatial-decay profiles $c(r)$ of the linear response of a glass to local force dipoles, defined and explained in detail in Appendix~\ref{xi_g_and_omega_g_appendix}. Figures~\ref{fig:all_core_length_observables}(a)-(c) show the decay functions $c(r)$, while Figs.~\ref{fig:all_core_length_observables}(d)-(f) show the products $r^{6}c(r)$, with the aim of visualizing the length $\xi_g$ beyond which the continuum scaling $c(r)\!\sim\!r^{-6}$ starts to hold. As demonstrated in Appendix~\ref{xi_g_and_omega_g_appendix}, we define the length $\xi_g$ as the location of the maximum of the products $r^{6}c(r)$, marked by the vertical dashed line in Figs.~\ref{fig:all_core_length_observables}(d)-(f), for the largest and smallest value of the relevant control parameter.

Our estimations of the length $\xi_g$ are shown in Fig.~\ref{fig:all_core_length_observables}(g)-(i); we observe that $\xi_g$ depends only weakly on $q$ in the QSS system, while it appears to depend strongly on the respective control parameters of the DSS and CSS systems, varying by up to a factor of 2. This variation is large compared to the one that stems from very deep supercooling of equilibrium parent configurations: in Ref.~\cite{pinching_pnas} a decrease of approximately 40\% was observed between $\xi_g$ measured for glasses quenched from high parents temperatures, and that measured in glasses quenched from very deeply supercooled states.

We next consider the effect of glass stickiness on the characteristic frequency and number density of QLMs.


\subsection{Quasilocalized modes' characteristic frequency and number density}
\label{sec:omega_g_and_calN}
In parallel to the decrease of QLMs' core size upon thermal annealing or deep supercooling, the characteristic frequency of QLMs has been shown to increase, by up to a factor of $\approx\!2$ \cite{modes_prl_2016, protocol_prerc,cge_paper,pinching_pnas}. It is therefore of interest to examine how the strength of attractive interactions of our model glasses affects their embedded QLMs' characteristic frequencies.

With the length $\xi_g$ in hand (see Fig.~\ref{fig:all_core_length_observables}), we are in position to estimate the characteristic frequency $\omega_{g}$ of QLMs in our different glass ensembles via the relation 
\begin{equation}\label{eq:omega_g}
    \omega_{g}=2\pi \frac{c_{s}}{\xi_{g}}\,,
\end{equation}
confirmed recently in Ref.~\cite{pinching_pnas}, but proposed in a similar form earlier \cite{sokolov_boson_peak_scale}. Our results are reported in Fig.~\ref{fig:wg_N_QLE}(a)-(c); we find that QLMs in the QSS model tend to not stiffen in any appreciable manner upon changing the exponent $q$. In contrast, the DSS and CSS systems show a substantial increase in the characteristic frequency $\omega_g$, by a factor of approximately $2$, by increasing glass stickiness.

Finally, thermal annealing processes have also been shown to deplete QLMs \cite{cge_paper,pinching_pnas,corrado_dipole_statistics_2020}, thus it is interesting to ask how the degree of glass stickiness affects the total number of QLMs in our glasses. Following Refs.~\cite{cge_paper,pinching_pnas}, we estimate the number density of QLMs for our different glass ensembles via
\begin{equation}\label{eq:cal_N}
    {\cal{N}}\!=\!A_g\omega_g^{5}\,. 
\end{equation}

Our results are displayed in Fig.~\ref{fig:wg_N_QLE}(d)-(f). As expected from the weak $q$-dependence of both $A_g$ and $\xi_g$ (and thus of $\omega_g$) in the QSS system, QLMs show no depletion in that system. The contrasting result for the DSS and CSS systems is rather interesting; Figs.~\ref{fig:wg_N_QLE}(e)-(f) show that the number of QLMs in these system are \emph{not} depleted by increasing glass stickiness, with the exception of $r_c\!=\!1.1$, which features an anomalously dilute population of QLMs: ${\cal N}\!\sim\!{\cal O}(10^{-2})$. This means that most of the change we find in $A_g$ in these models (see Fig.~\ref{fig:prefactor}) stems from the stiffening of~QLMs with increasing glass stickiness --- reflected by the variation of $\omega_g$ as shown in Fig.~\ref{fig:wg_N_QLE}a-c --- rather than from their depletion, as discussed at length in Refs.~\cite{cge_paper,pinching_pnas}. In the most extreme case, namely the CSS glasses with $r_c\!=\!1.1$, the change in the prefactor $A_g$ cannot be fully accounted for by the stiffening of $\omega_g$, indicating that QLMs are then \emph{depleted}.

One more interesting observation should be mentioned; up to some noticeable noise in ${\cal N}$ (stemming from uncertainties, primarily in the extraction of  $A_g$, but also of $\xi_g$, cf.~Eqs.~(\ref{eq:omega_g}) and (\ref{eq:cal_N})), the value of ${\cal N}\!\approx\!0.3$ appears to be quasi-universal \cite{footnote_ncal}, for all of the ensembles studied here, with the aforementioned exception of the CSS system with $r_c=1.1$, the most stable amongst all systems studied. Understanding the precise mechanism that leads not only to the observed substantial stiffening, but also to the depletion of QLMs, is left for future studies. 

We note finally that ${\cal N}$ is not expected to remain bounded from above near the unjamming transition; in Refs.~\cite{ikeda_pnas,atsushi_core_size_pre} it was shown that disordered packings of harmonic spheres confined at pressure $p$ feature $A_g\!\sim\!p^{-4}$. Associating $\omega_g$ with the characteristic frequency scale $\omega^\star$ of anomalous modes \cite{matthieu_PRE_2005,Silbert_prl_2005}, one expects $\omega_g\!\sim\!\omega^\star\!\sim\!\sqrt{p}$, then Eq.~\eqref{eq:cal_N} implies that ${\cal N}\!\sim\!p^{-3/2}$, i.e.~it diverges near unjamming. Further work is needed to establish the relevance of Eq.~\eqref{eq:cal_N} to unjamming.


\section{Interpretations, insights and propositions}
\label{sect:discussion}

Using two simple computer glass models in which the strength of attractive interactions can be tuned, we demonstrated that glasses' macro- and microscopic elastic properties, mechanical disorder and stability, can depend dramatically on the strength \emph{and} on the functional form of pairwise attractive interactions. In what follows we hold extensive discussions about several points of interest.

\subsection{How do features of attractive interaction potentials affect glass stability?}
\label{sec:discussion_freq_frac}

Why does increasing glass stickiness lead to the mechanical stabilization of glasses in some cases (cf.~DSS and CSS glasses), but not in others (cf.~QSS glasses)? Intuitively, it is not surprising that attractive forces themselves stabilize mechanical structures, be them ordered \cite{Rossi2011, Sun2013} or disordered \cite{brian_sticky_prl}. In the absence of attractive forces or a sufficient confining pressure, solidity itself would be lost, as happens near the unjamming point \cite{ohern2003,liu_review,van_hecke_review}. To better understand how the observed attraction-induced stabilization comes about, we next examine how the energies of quasilocalized, nonphononic low-frequency modes are comprised out of different contributions of the linear stability (Hessian) operator of our computer glasses. 

In models employing radially-symmetric pairwise potentials, the Hessian of the potential energy can be split into four terms \cite{matthieu_PRE_2005,Silbert_prl_2005,mw_prl_2007_kablj_unjamming,eric_boson_peak_emt,Silbert_pre_2016_jamming,inst_note}:
\begin{eqnarray}
    \calBold{M} & = & \sum\limits_{\varphi_{ij}''>0}\varphi_{ij}''\frac{\partial r_{ij}}{\partial\xv}\otimes\frac{\partial r_{ij}}{\partial\xv} + \sum\limits_{\varphi_{ij}''<0}\varphi_{ij}''\frac{\partial r_{ij}}{\partial\xv}\otimes\frac{\partial r_{ij}}{\partial\xv} \nonumber \\
    & &\ \  + \sum\limits_{\varphi_{ij}'>0}\varphi'_{ij}\frac{\partial^2r_{ij}}{\partial\xv\partial\xv} + \sum\limits_{\varphi_{ij}'<0}\varphi'_{ij}\frac{\partial^2r_{ij}}{\partial\xv\partial\xv} \\
    & \equiv & \ \  \calBold{M}''_+ +  \calBold{M}''_- + \calBold{M}'_+ + \calBold{M}'_- \,, \label{eq:argument}
\end{eqnarray}
where $\calBold{M}''_+,\calBold{M}'_+$ are positive definite, and $\calBold{M}''_-,\calBold{M}'_-$ are negative definite. With the above decomposition of $\calBold{M}$, the energy $\omega^2\!=\!\uv\cdot\!\calBold{M}\!\cdot\uv>\!0$ associated with any given, translation-free mode $\uv$ can be written as
\begin{equation}\label{eq:frequency_decomposition}
    \omega^2 = k_+ + k_- + f_+ + f_- \,,
\end{equation}
where 
\begin{eqnarray}
k_+ & \equiv & \uv\cdot\calBold{M}''_+\cdot\uv\,,\quad\quad k_-  \equiv  \uv\cdot\calBold{M}''_-\cdot\uv\,, \nonumber \\
f_+ & \equiv & \uv\cdot\calBold{M}'_+\cdot\uv\,,\quad\quad f_-  \equiv  \uv\cdot\calBold{M}'_-\cdot\uv\,, \nonumber  
\end{eqnarray}
and we note that $k_+,k_-,f_+,f_-$ all have units of energy/length$^2$, assuming $\uv$ is normalized and thus dimensionless.

In widely-studied purely-repulsive models such as inverse-power-law or Hertzian spheres (see e.g., Ref~\cite{boring_paper}), $f_+\!=\!k_-\!=0$ identically, by construction. It is known that delicate cancellations between the remaining $k_+\!>\!0$ and $f_-\!<\!0$ terms in Eq.~(\ref{eq:frequency_decomposition})  to give rise to the low frequencies featured by quasilocalized modes (QLMs) \cite{inst_note}, and by other classes of nonphononic modes \cite{eric_boson_peak_emt,Silbert_pre_2016_jamming,atsushi_high_dimension_dos} in systems residing near the unjamming transition \cite{ohern2003,liu_review,van_hecke_review}. In the latter case, this delicate cancellation is referred to as `marginal stability' \cite{mw_mm_marginal_stability_2015}, and allows one to deduce scaling relations between the mean coordination and elastic moduli \cite{matthieu_PRE_2005,Silbert_prl_2005}.

Introducing attractive interaction terms on top of repulsive ones gives rise to nonzero stabilizing $f_+$ and destabilizing $k_-$ contributions to the energy $\omega^2$ \cite{footnote2}, cf.~Eq.~(\ref{eq:frequency_decomposition}). It is the interplay between these different contributions that determines glass stability, i.e.~its featured abundance of soft QLMs as reflected by the dimensionless prefactor~$A_g\omega_0^5$. To assess the \emph{relative} contributions of the different terms to the energy of a mode, we consider the rescaled energies $k_+/\omega_0^2, k_-/\omega_0^2, f_+/\omega_0^2$ and $f_-/\omega_0^2$, and recall that $\omega_0^2\!\equiv c_s^2/a_0^2$, with $c_s$ denoting the speed of shear waves, and $a_0$ denoting an interparticle distance. We focus in particular on the QSS system with $q\!=\!20$ (see `glass 1' in Fig.~\ref{fig:fig1}) and the DSS system with $\rho\!=\!0.55$ (see `glass 2' in Fig.~\ref{fig:fig1}), whose pairwise potentials, Poisson's ratios, and probability distributions of low-frequency vibrational modes were reported in Fig.~\ref{fig:fig1}. The attractive parts of the respective pairwise potentials of these models have roughly similar forms, however the emergent elastic properties of the resulting glasses of these models are found to be quite different, see further details in Sec.~\ref{sec:intro}.

\begin{figure}[ht!]
\centering
  \includegraphics[width=1.0\linewidth]{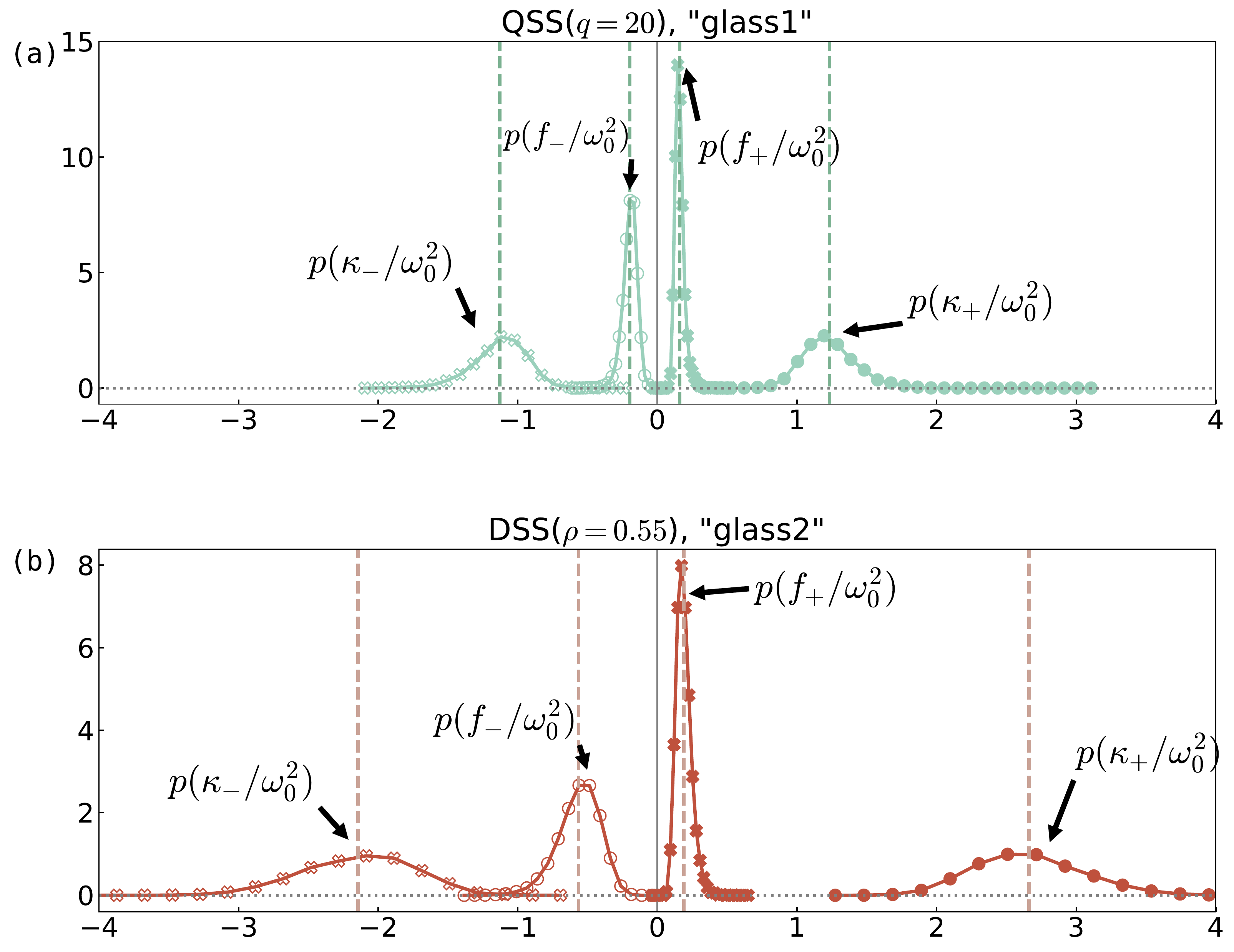}
\caption{\footnotesize Distributions of relative contributions to QLMs' energies (see text for precise definitions), measured for (a) the QSS system with $q\!=\!20$ and (b) for the DSS system with $\rho\!=\!0.55$. The vertical dashed lines represent the means of each relative contribution.
\label{fig:freq_frac_distributions}}
\end{figure}

To the aim of revealing the origin of the aforementioned differences between the QSS, $q\!=\!20$ and DSS, $\rho\!=\!0.55$ glasses, we examine the distributions of the rescaled energy contributions as defined above, extracted for ensembles of soft, quasilocalized modes, calculated as explained in Appendix~\ref{app:eigenstrains} (one soft QLM per glass). The results are shown in Fig.~\ref{fig:freq_frac_distributions}. We make three key observations regarding these distributions.

The first observation is that, despite that the frequencies of the QLMs we calculated (as described in Appendix~\ref{app:eigenstrains}) in both the QSS and DSS systems span over half a decade over the frequency axis, the distributions of the \emph{dominant} contributions to the modes' energies feature relative widths of order unity. Crucially, we find that these relative widths are $N$-independent, conditioned that the frequencies of the modes calculated reside well within the $\sim\!\omega^4$ scaling regime of the nonphononic vDOS. This indicates that while the softening mechanism of QLMs can be particular to each glass model, it remains predominantly independent of the modes' frequencies within a given glass model, at low frequencies.

The second observation is that, in our sticky-spheres models, the dominant softening mechanism of QLMs comes from the exploitation of large \emph{negative pairwise stiffnesses}, namely from the $k_-$ contribution to the total energy $\omega^2$. This stands in essential contrast with the softening mechanism of QLMs in purely repulsive systems, for which $|f_-|\!\sim\!k_+$ \cite{inst_note}. In addition, and perhaps counter-intuitively, the stabilizing effect of attractive \emph{forces} is relatively very small, approximately an order of magnitude smaller than the contribution of positive, stabilizing stiffnesses.

Finally, the third observation is that the approximate symmetry around the origin between the dominant pair $p(k_+/\omega_0^2)$ and $p(k_-/\omega_0^2)$, as seen in the QSS system, is violated in the DSS system: both $p(k_+/\omega_0^2)$ and $p(k_-/\omega_0^2)$ feature means (marked by dashed vertical lines in Fig.~\ref{fig:freq_frac_distributions}) larger in amplitude and in bias towards positive energy contributions, compared to the same distributions in the QSS system. We note, importantly, that as $N\!\to\!\infty$, the sum of the means of the 4 dimensionless contributions to modes' energies --- which represents the mean minimal QLM frequency in a system of $N$ particles --- is expected to vanish as $\sim\!N^{-2/5}$~\cite{modes_prl_2016,cge_paper}, so long as the glasses analyzed feature gapless $\sim\!\omega^4$ nonphononic spectra. However, the key observation here is that, in the more stable glasses (DSS with $\rho\!=\!0.55$), the destabilizing effect of the $k_-$ term is relatively weaker, suggesting that the population of interactions featuring large, negative stiffnesses in the DSS glasses is depleted, compared to its size in the QSS glasses. We speculate therefore that the relative mechanical instability of the QSS glasses compared to the DSS glasses, as shown in Fig.~\ref{fig:fig1}, stems from the presence of a larger \emph{population} of negative-stiffness interactions in the former.

\begin{figure}[ht!]
\centering
  \includegraphics[width=1.0\linewidth]{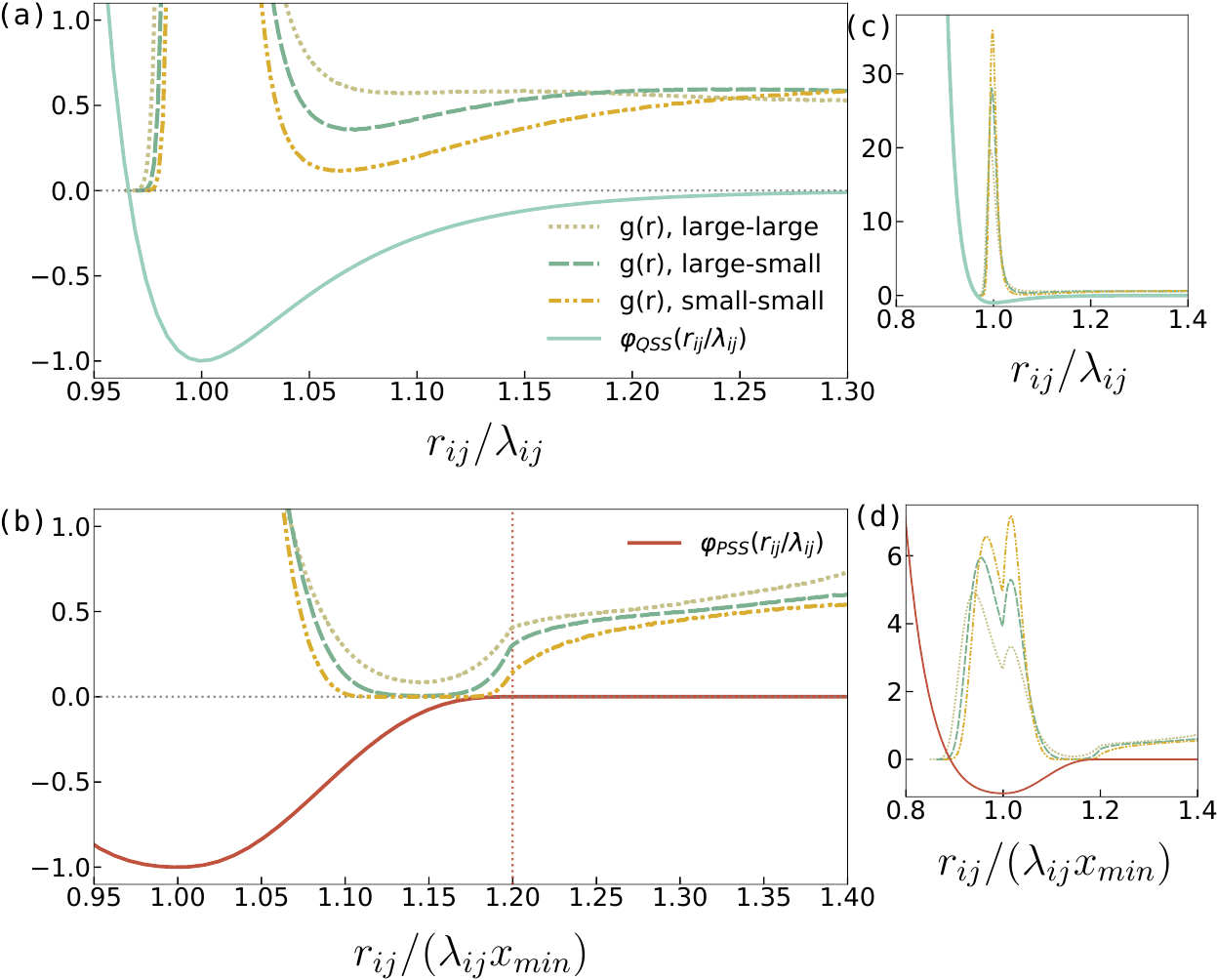}
\caption{\footnotesize The pairwise potentials $\varphi_{\mbox{\tiny QSS}}$ (a) and $\varphi_{\mbox{\tiny PSS}}$ (b) are superimposed with their resulting glasses' radial distribution functions $g(r)$, calculated for different pair types as detailed in the legends, and plotted against the dimensionless distance $r_{ij}/\lambda_{ij}$. The key observation here is that large, negative stiffnesses in the pairwise potential lead to a depletion of the population of interactions featuring those stiffnesses, which can affect, in turn, QLMs' stiffnesses and thus glass stability. The vertical dashed line in panel (b) marks the cutoff $r_{c}=1.2$ used for the DSS system. Panels (c) and (d) are zoomed-out representations of panels (a) and (b), respectively. 
\label{fig:potential_overlapped_gr}}
\end{figure}

\begin{figure}[h!]
\centering
  \includegraphics[width=1.0\linewidth]{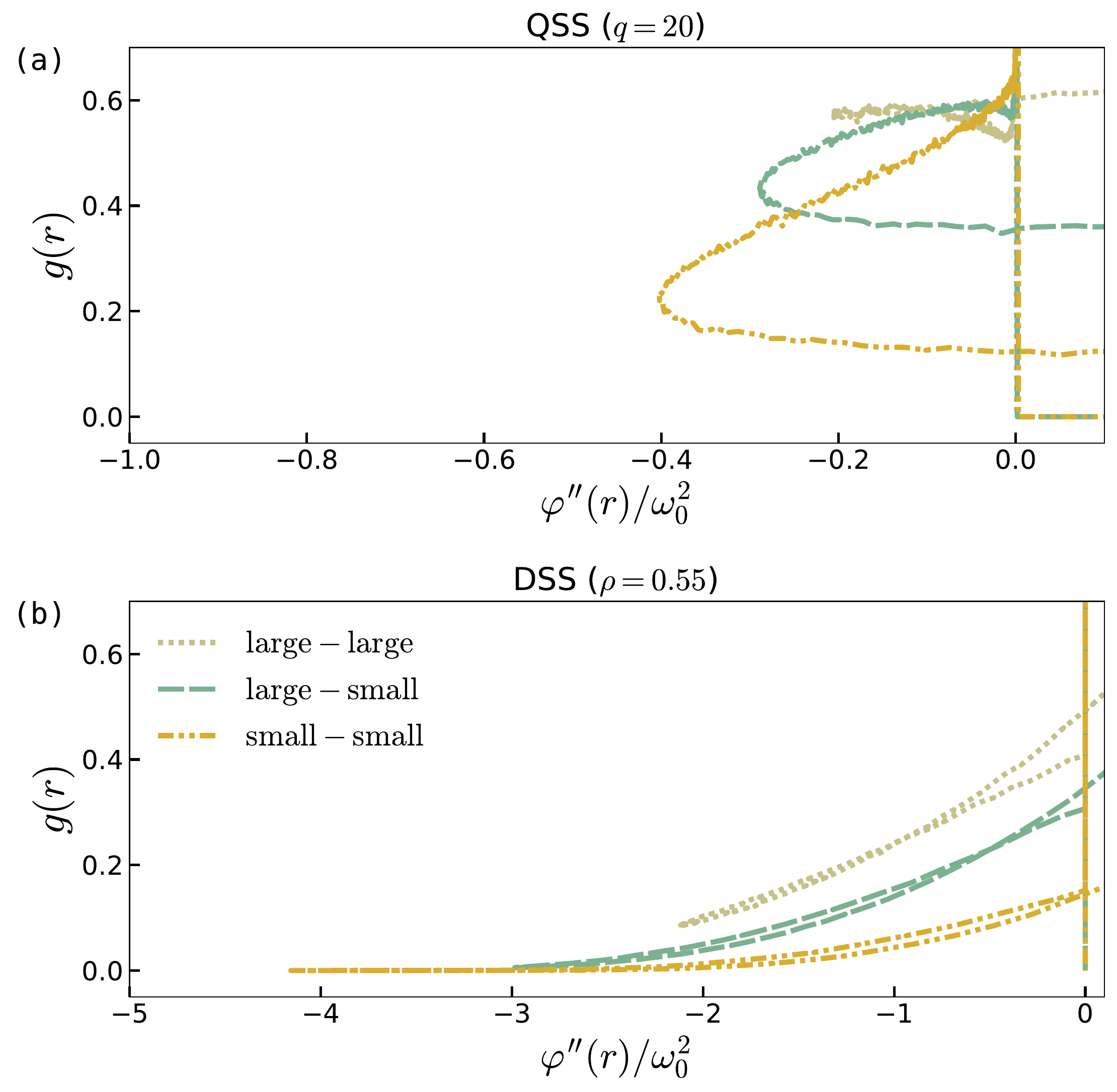}
\caption{\footnotesize Parametric plots of the radial distribution $g(r)$ vs.~the pairwise stiffness $\varphi''(r)$ for (a) the QSS system with $q\!=\!20$ and (b) the DSS system with $\rho\!=\!0.55$. Here we consider only distances $r$ for which $\varphi''(r)\!<\!0$. We clearly see that the population of interactions featuring large negative stiffnesses becomes smaller at more negative stiffnesses.
\label{fig:gr_vs_stiff}}
\end{figure}

The depletion process suggested above is validated in Fig.~\ref{fig:potential_overlapped_gr}, which shows that indeed as the pairwise potential features larger, more negative stiffnesses, the population of interactions in the resulting glasses that feature those negative stiffnesses is depleted. Their depletion --- as indicated by the reduction of the radial distribution function $g(r)$ --- is stronger in the DSS system (Fig.~\ref{fig:potential_overlapped_gr} (b)), where `small'-`small' interactions with the largest negative stiffnesses are, as a result, entirely absent. In contrast, we see a significantly weaker depletion of the `small'-`small' and `large-small' negative-stiffness interactions in the QSS system (compare Fig.~\ref{fig:potential_overlapped_gr} (a) and Fig.~\ref{fig:potential_overlapped_gr}(b)), and almost no negative-stiffness-induced depletion in the population of `large-large' interactions of the QSS system. To clarify this point further, in Fig.~\ref{fig:gr_vs_stiff} we parametrically-plot the pair correlation function $g(r)$ against the (negative) dimensionless stiffness $\varphi''(r)/\omega_0^2$ for the QSS and DSS glasses, in the range of distances $r$ for which $\varphi''(r)<0$. This representation further sharpens the depletion picture discussed above.

\begin{figure*}[ht!]
\centering
  \includegraphics[width=0.65\linewidth]{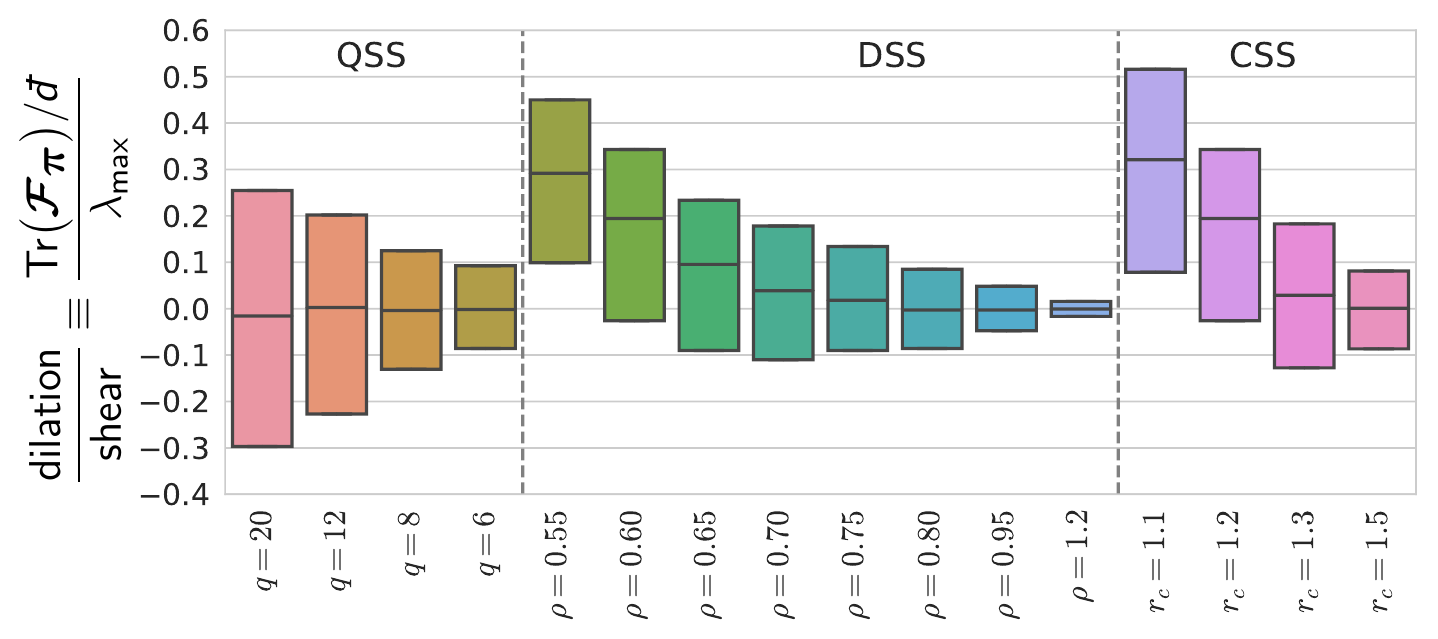}
\caption{\footnotesize The bars cover the second and third quartiles of the ratio of shear to dilational strain coupling of QLMs, and the middle horizontal line represents these ratios' averages, see Appendix~\ref{app:eigenstrains} for precise definitions. 
\label{fig:dilation_shear}}
\end{figure*}

To summarize, according to the physical picture proposed here, negative stiffnesses featured by attractive pairwise potentials can serve as the main softening mechanism of low-frequency nonphononic modes, as we showed for the QSS, $q\!=\!20$ glasses. However, when those negative stiffnesses are made very large (compared to characteristic positive interaction stiffness scales), as seen in the DSS, $\rho\!=\!0.55$ glasses, the population of interactions that possess those large negative stiffnesses becomes depleted, resulting in the stabilization of the glass, and the reduction of its shear modulus fluctuations, via the stiffening of its typical QLMs' frequencies.

\subsection{Fluctuations of elastic moduli}
\label{sect:elastic_moduli_fluctuation}

Another interesting trend we have identified is the opposite variations with glass stickiness of relative fluctuations of shear and bulk elastic moduli, the latter are captured by $\chi_{_G}$ and $\chi_{_K}$, respectively, and reported in Fig.~\ref{fig:chi}. In particular, we find that $\chi_{_G}$ decreases, while $\chi_{_K}$ increases, upon increasing glass stickiness -- with the interesting exception of $\chi_{_G}$ in the QSS glasses, that appears to be invariant to changing glass stickiness.

What is the origin of these opposite trends? We proposed above that the stiffening and depletion of QLMs should lead to the \emph{reduction} in moduli fluctuations, due to their diminished effect on moduli's respective nonaffine terms \cite{Silbert_pre_2016_jamming,stefanz_pre}. The latter are shown below to manifest the influence of soft modes on elastic moduli and their fluctuations. This relation between soft modes and elastic moduli can explain, on a qualitative level, the trends seen in $\chi_{_G}$ (see Fig.~\ref{fig:chi}b,c), which decreases substantially upon increasing glass stickiness in the DSS and CSS systems.

One more physical factor that may control the effect of QLMs' properties on elastic moduli fluctuations of glasses can be identified by writing the nonaffine term of any elastic moduli $E$ as \cite{lutsko}
\begin{equation}\label{eq:nonaffine_term}
    E_{\mbox{\scriptsize na}} \equiv \frac{\frac{\partial^2U}{\partial\epsilon\partial\xv}\cdot\calBold{M}^{-1}\cdot\frac{\partial^2U}{\partial\xv\partial\epsilon}}{V} = \frac{1}{V}\sum_\ell\frac{\big(\mathBold{\psi}^{(\ell)}\!\cdot\!\frac{\partial^2U}{\partial\xv\partial\epsilon}\big)^2}{\omega_\ell^2}\,,
\end{equation}
where $\epsilon$ represents a strain parameter (cf.~the shear and dilational strain parameters $\gamma$ and $\eta$ of Eqs.~(\ref{shear_transformation_matrix}) and (\ref{dilation_transformation_matrix}) in Appendix~\ref{appx:macro}, respectively), and $\mathBold{\psi}^{(\ell)}$ is the eigenmode of $\calBold{M}$ pertaining to the eigenvalue $\omega_\ell^2$. This expression clarifies that not only do the statistics of soft modes affect the typical values and fluctuations of elastic moduli, but so does the strength of their coupling to different deformation geometries.

Recently, a set of tools to extract the core properties of QLMs, including their deformation-coupling, was developed in Ref.~\cite{avraham_arXiv}. Here we employ those tools to probe the way QLMs in our different computer glasses couple to external deformations; see a detailed explanation about the QLMs, their deformation-coupling definitions, and calculations in Appendix~\ref{app:eigenstrains}. The analysis determines the ratio of a QLM's dilatational strain coupling to shear strain coupling, referred to here as simply `dilation/shear'. This ratio is calculated for a single QLM extracted for each member (i.e.~one QLM per glassy sample) of our different sets of glass ensembles.

The results of these calculations are shown in Fig.~\ref{fig:dilation_shear}; we find that glass stickiness has a pronounced effect on the \emph{mean} dilation vs.~shear coupling of QLMs (middle line on bars), which increases dramatically with increasing glass stickiness in the DSS and CSS ensembles. This is an interesting observation on its own right -- that the form of the interaction potential of a glass has a strong and systematic effect on the structural and mechanical properties of that glass's QLMs. To the best of our knowledge, this observation has not been made in previous literature; an extensive investigation of these effects and their importance is left for future studies. 

In the QSS system the mean dilation vs.~shear coupling remains vanishing small for all $q$'s (see leftmost bars in Fig.~\ref{fig:dilation_shear}); however, the \emph{fluctuations} of these coupling ratios increase dramatically with increasing $q$. As can be seen from Eq.~(\ref{eq:nonaffine_term}), the square of the couplings $\big(\mathBold{\psi}^{(\ell)}\!\cdot\!\frac{\partial^2U}{\partial\xv\partial\epsilon}\big)^2$ enter the expression for the nonaffine modulus, implying that fluctuations in the dilation to shear coupling ratio would be echoed by fluctuations of the associated elastic moduli, regardless of the former's fluctuations sign. This mechanism is suggested to qualitatively explain the behavior of $\chi_{_K}$ as seen in Fig.~\ref{fig:chi}d. Curiously, $\chi_{_K}$ and the fluctuations in the shear to dilation coupling ratios discussed here are the only observables considered in our work that show a substantial dependence on the exponent $q$ in the QSS glasses. 

We note finally that similar ideas regarding QLMs' deformation couplings were put forward in the context of the elastic properties of metallic glasses in Ref.~\cite{experimental_inannealability}. In that work a distinction is made between `shear softening' and `pressure softening' of `local soft regions', which echoes some of the differences we showed here between QLMs' properties in non-sticky and sticky glasses, respectively. We speculate that our framework of QLMs, their abundance, and their deformation-coupling properties, are concrete micromechanical realizations of the `shear softening' and `pressure softening' concepts put forward in Ref.~\cite{experimental_inannealability}.

\subsection{Glasses' STZ size vs.~Poisson's ratio}


Plastic flow in structural glasses is known to proceed via dissipative, localized rearrangements of a few tens or hundreds of particles \cite{argon_st,spaepen_1977}. The precursors of those rearrangements --- called Shear Transformation Zones (STZs) ---, have been recently shown using computer simulations to correspond to a subset of soft QLMs \cite{phm_arXiv}. This allows us to meaningfully compare between QLMs' core size measured in our computer glasses, to the sizes of shear transformation zones (STZs) measured in the experiments on bulk metallic glasses reported in Ref.~\cite{stz_experimental}. To this aim we estimated the STZ size as $c_{\mbox{\tiny STZ}}\,\xi_g^3$, 
with $c_{\mbox{\tiny STZ}}$ a constant of order unity.

\begin{figure}[h!]
\centering
  \includegraphics[width=1.0\linewidth]{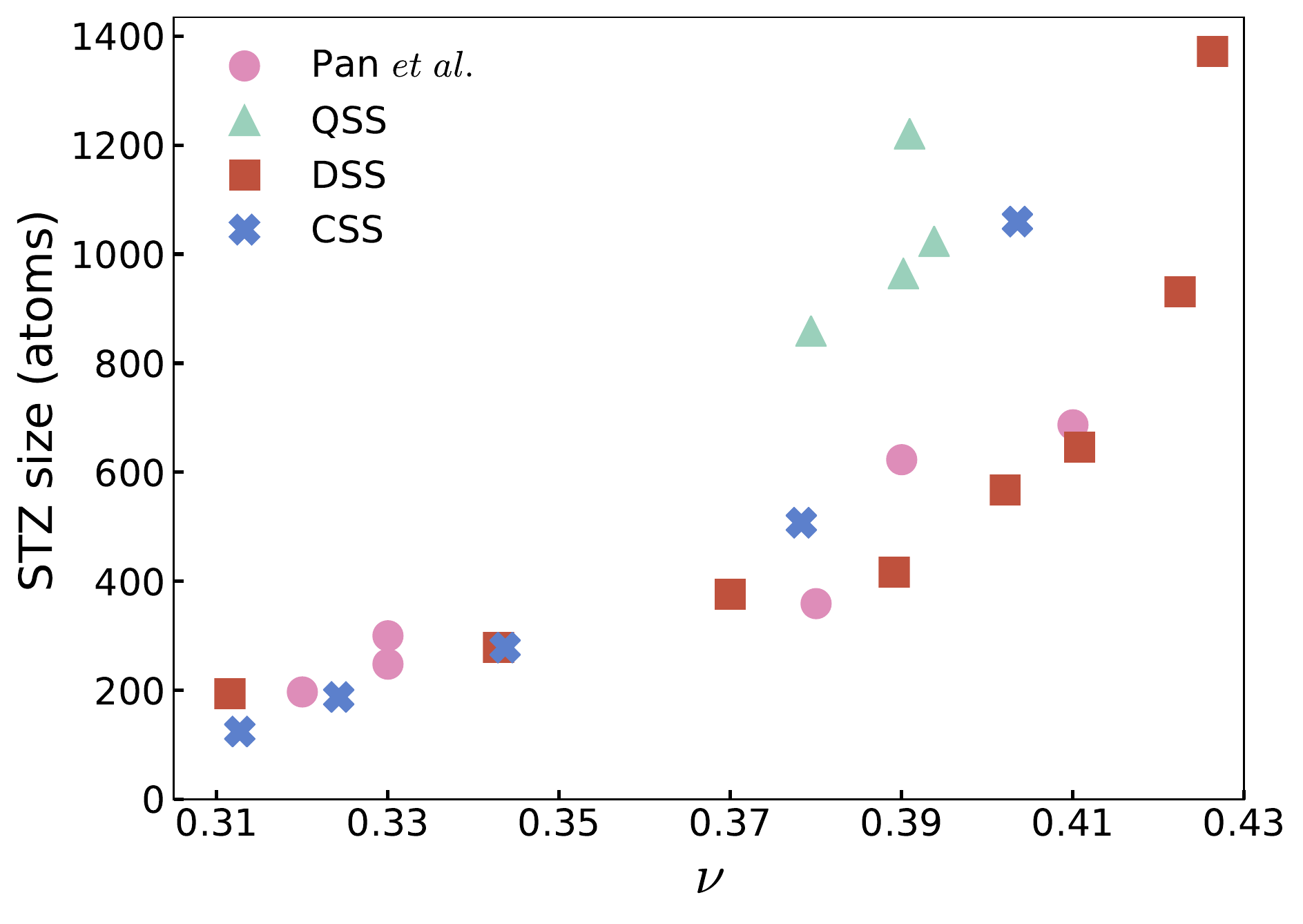}
\caption{\footnotesize Comparison between simulational and experimental \cite{stz_experimental} size of STZs, and their dependence on the Poisson's ratio. We estimate the STZ size for our computer glasses to be proportional to QLMs' core size $c_{\mbox{\tiny STZ}}\,\xi_g^3$, and plot it against those glasses' Poisson's ratio (we used $c_{\mbox{\tiny STZ}}\!=\!2$). STZ sizes were estimated experimentally in various laboratory metallic glasses in Pan~\emph{et al}.~\cite{stz_experimental}. 
\label{fig:stz}}
\end{figure}

The results are shown in Fig.~\ref{fig:stz}, where we plot the estimated STZ size against the Poisson's ratio of the glasses that host those STZs. The agreement is satisfying, both in magnitude as well as in trend; in the laboratory metallic glasses, and in our computer glasses, the size of STZs grows upon increasing the hosting glass's Poisson's ratio. At high Poisson's ratios the scatter in STZ sizes increases, indicating that the correspondence between STZ size and Poisson's ratio is not one-to-one.


\subsection{Is the Poisson's ratio an indicator of a glass's susceptibility to plastic flow?}

We next comment on the possible relation between a glass's Poisson's ratio $\nu$ and its degree of ductility or brittleness, as suggested in previous work, e.g., Refs~\cite{brittleness_BMG,stz_experimental,shi_intrinsic_ductility}. We first note that in Refs.~\cite{itamar_brittle_to_ductile_pre_2011,smarajit_ductile_brittle_soft_matter_2016} it was shown that CSS glasses with small $r_c$ feature brittle failure, while larger $r_c$ glasses feature ductile failure. Since we find that the prefactor $A_g$ depends strongly on $r_c$ --- while ${\cal N}$ is mostly constant ( cf.~Fig.~\ref{fig:wg_N_QLE}) --- we conclude that it is $A_g$ that facilitates or inhibits plasticity via the abundance or sparsity of STZs, respectively. 

\begin{figure}[ht!]
\centering
  \includegraphics[width=1.00\linewidth]{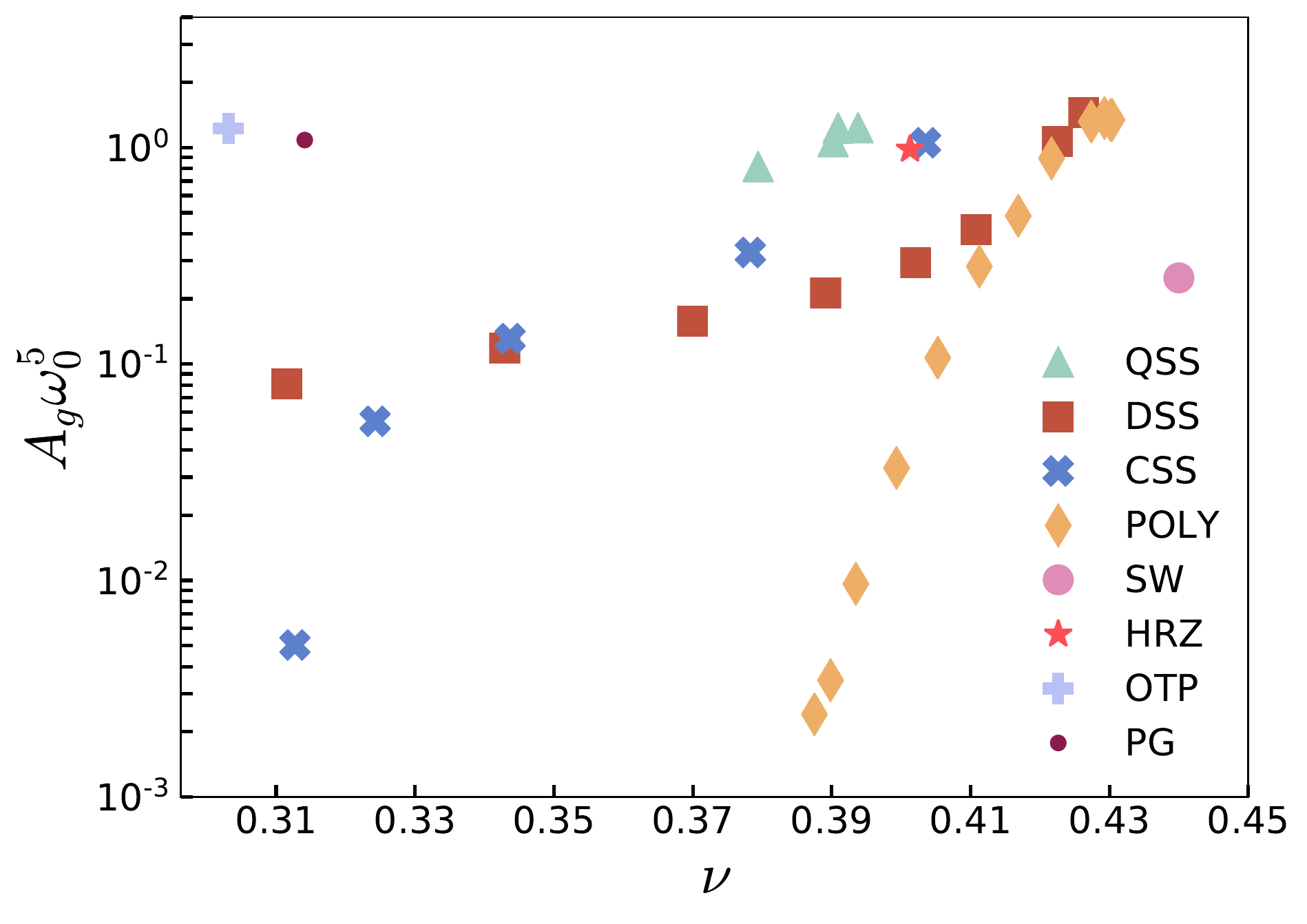}
\caption{\footnotesize Dimensionless prefactors $A_g\omega_0^5$ of the nonphononic vDOS (see Eq.~(\ref{d_of_omega}) and Fig.~\ref{fig:prefactor}), plotted against the Poisson's ratio $\nu$, see text for discussion.
\label{fig:ag_vs_nu}}
\end{figure}

For these reasons, we plot in Fig.~\ref{fig:ag_vs_nu} the dimensionless prefactors $A_g\omega_0^5$ of our glass ensembles, against those glasses' mean Poisson's ratios $\nu$. In addition, we measure and plot in Fig.~\ref{fig:ag_vs_nu} the same observables for polydisperse, inverse-power-law soft-sphere computer glasses quenched from a broad range of parent temperatures (`POLY' in the figure legend, see model details and parameters in Ref.~\cite{pinching_pnas}), for Hertzian binary-mixture soft-sphere computer glasses (HRZ in the figure legend, see model details in Ref.~\cite{modes_prl_2020}), for the Stillinger-Weber tetrahedral-network glass-former \cite{Stillinger_Weber} (SW in the figure legend), for the Orthophenyl molecular-glass former (OTP in the figure legend, for details on the model see Ref.~\cite{modes_prl_2020}), and for a polymeric glass (PG in the figure legend, see details also in Ref.~\cite{modes_prl_2020}). We find that while a single model's Poisson's ratio $\nu$ seems to form a monotonically increasing function with that model's prefactor $A_g$, there is no deep, overarching relation between these observables, as indicated by the huge spread of $A_g$ seen at almost all $\nu$ values, once  different model systems are considered. Similar statements were previously made in Refs.~\cite{eran_fracture_toughness_prl_2012,KUMAR2011585,jan_2020, AIHEMAITI2020106834}.

\subsection{Micro- vs.~macroscopic quantifiers of disorder}

The satisfying agreement we find between the experimental and simulational STZ sizes, and their correlated variation with the Poisson's ratio $\nu$, is an interesting connection between micro- and macroelastic  observables. It suggests a possible connection between macroscopic and microscopic quantifiers of disorder, which we test next. 

\begin{figure}[ht!]
\centering
  \includegraphics[width=0.95\linewidth]{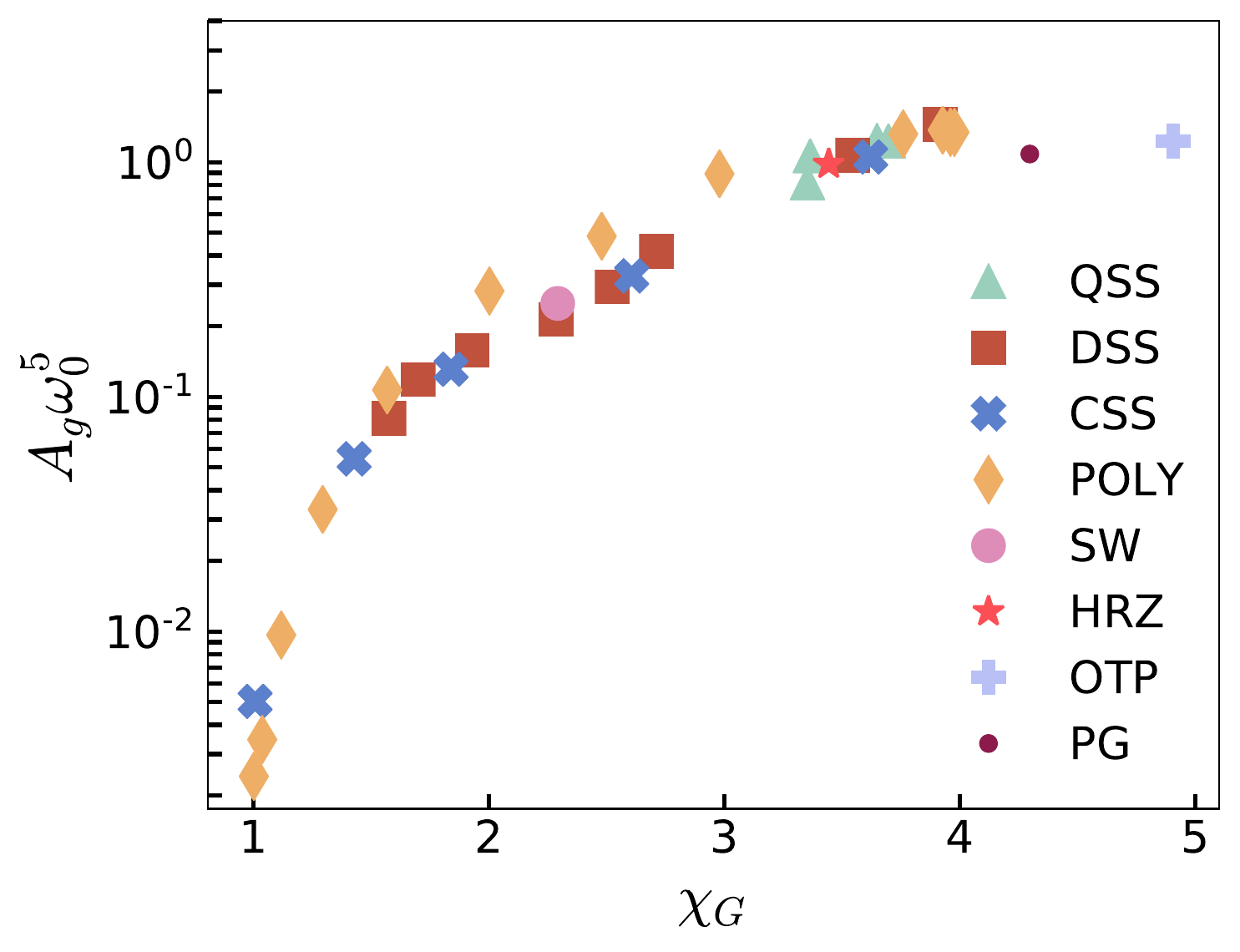}
\caption{\footnotesize  Dimensionless prefactors $A_g\omega_0^5$ of the nonphononic vDOS (see Eq.~(\ref{d_of_omega}) and Fig.~\ref{fig:prefactor} above), plotted against the dimensionless, $N$-independent quantifier $\chi_{_G}$ of sample-to-sample shear modulus fluctuations, for the sticky-sphere glass ensembles, and for 5 additional glass models: polydisperse power-law (POLY), Stillinger-Weber (SW), Hertzian soft spheres (HRZ), Orthophenyl (OTP), and polymeric (PG) see text for details.
\label{fig:ag_vs_chig}}
\end{figure}

In Fig.~\ref{fig:ag_vs_chig} we plot the dimensionless prefactor $A_g\omega_0^5$ of the vDOS (reported in Fig.~\ref{fig:prefactor} ) against the dimensionless (and $N$-independent, see Eq.~(\ref{eq:chi})) measure $\chi_{_G}$ of sample-to-sample fluctuations of the shear modulus (reported in Fig.~\ref{fig:chi}). We also include data measured for the POLY, HRZ and SW glasses, see details about these models in the previous Subsection. Despite some measurable scatter, the good correlation is apparent and impressive, constituting a key result of our work, and another interesting link between microscopic mechanical fluctuations and macroscopic ones. 

\begin{figure}[ht!]
\centering
  \includegraphics[width=1.00\linewidth]{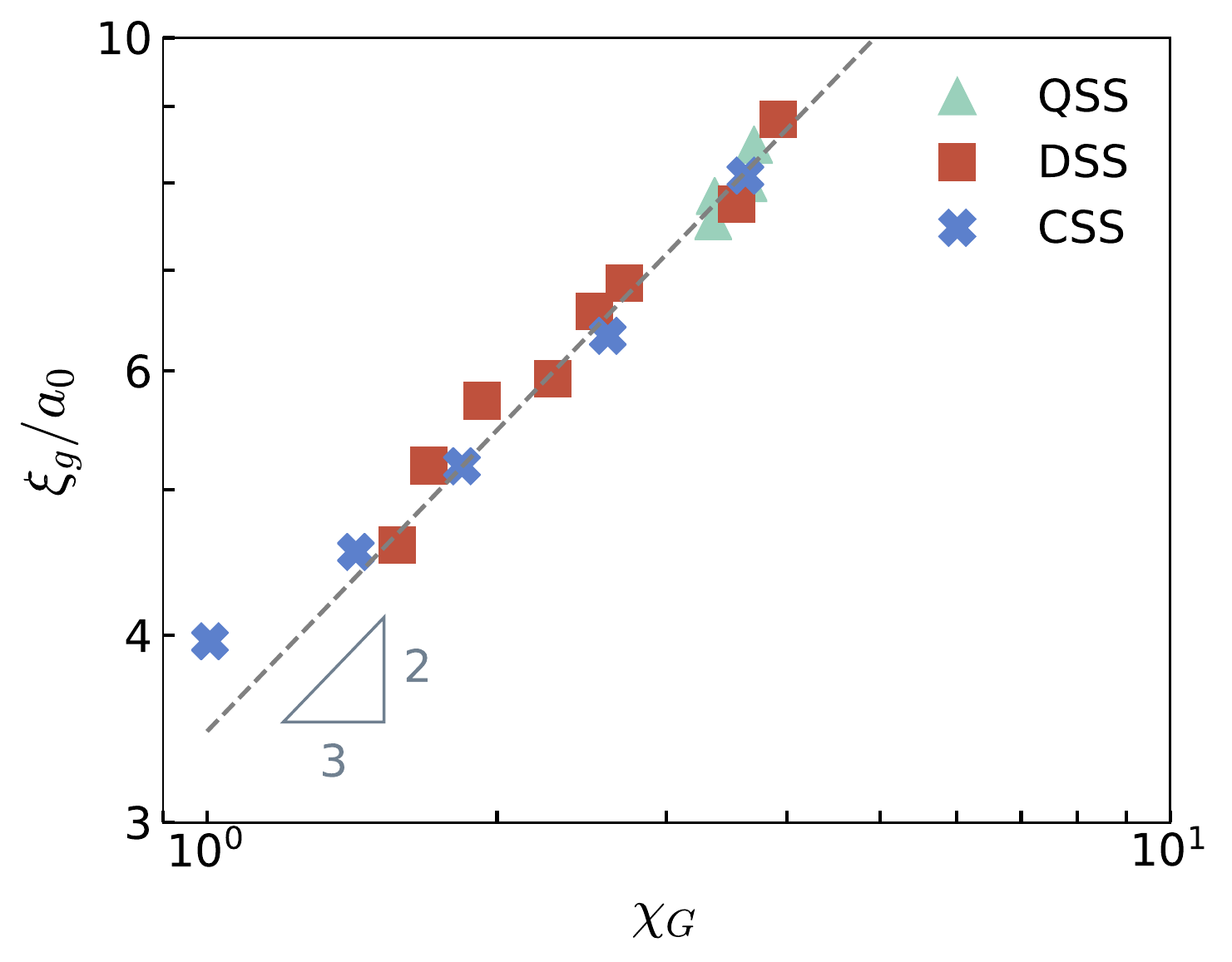}   
\caption{\footnotesize The crossover length $\xi_g$ (extracted as shown in Fig.~\ref{fig:length_extraction}), plotted against the sample-to-sample shear modulus fluctuations as quantified by $\chi_{_G}$. We find $\xi_g\!\sim\!\chi_{_G}^{2/3}$, for which scaling arguments are provided in the text.
\label{fig:xig_vs_chig}}
\end{figure}

We end this discussion section with one final observation, shown in Fig.~\ref{fig:xig_vs_chig}, where we plot the dimensionless length $\xi_g/a_0$ vs.~$\chi_{_G}$, for the QSS, DSS and CSS systems, on logarithmic scales. The dashed line corresponds to a $\xi_g/a_0\!\sim\!\chi_{_G}^{2/3}$ scaling, explained next.

Recall that QLMs characteristic frequency $\omega_g$ is related to the crossover length $\xi_g$ between micromechanical-fluctuations-dominated to continuum-elastic-like responses via $\omega_g/\omega_0\!\sim\!a_0/\xi_g$, as established in Ref.~\cite{pinching_pnas} and discussed in Sec.~\ref{sec:omega_g_and_calN} above. A similar crossover between heat transport via propagating phonons, to heat transport via `diffusons' \cite{diffusons1,diffusons2} occurs at the Ioffe-Regel frequency $\omega_{\mbox{\tiny IR}}$, defined implicitly via the frequency-dependent transverse wave attenuation rate $\Gamma(\omega)$ as $\Gamma(\omega_{\mbox{\tiny IR}})\!\sim\!\omega_{\mbox{\tiny IR}}$. According to Heterogeneous Elasticity Theory (HET) \cite{Schirmacher_2006, Schirmacher_prl_2007, Marruzzo2013}
\begin{equation}\label{eq:HET}
    \Gamma(\omega)\sim \chi_{_G}^2\omega^{\dbar+1}\,,
\end{equation}
at low wave frequencies $\omega$, as verified recently in two-dimension computer glasses in~\cite{scattering_prl_2020}, and recall that $\dbar$ denotes the dimension of space. Assuming that Eq.~(\ref{eq:HET}) persists up to $\omega_{\mbox{\tiny IR}}$, we expect that $\omega_{\mbox{\tiny IR}}\!\sim\!\chi_{_G}^{-2/\dbar}$. 

Given the above discussion, a reasonable proposition would be to to associate $\omega_g$ with $\omega_{\mbox{\tiny IR}}$ (see also Refs.~\cite{Monaco_prl_2006_boson_peak_ioffe_regel,Parshin_prb_2013_IR}), in which case we expect
\begin{equation}
    \xi_g/a_0 \sim \chi_{_G}^{2/\dbar}\,,
\end{equation}
as we indeed show in Fig.~\ref{fig:xig_vs_chig} for our computer glasses, forming yet another interesting and potentially-useful connection between micro- and macroscopic quantifiers of mechanical disorder.

\section{Summary of main Results}
\label{sect:summary}

In this work we have studied how increasing the strength of attractive interactions (referred to here as `glass stickiness') between the particles of simple computer glasses affect those glasses' micro- and macroscopic elastic properties, and their featured degree of mechanical disorder. Our main findings are the following:
\begin{enumerate}
\item The degree of mechanical disorder featured by glasses that were quickly quenched from high-$T_p$ liquid-configurations can vary substantially due to changes in their interparticle potential. In particular, we find that the relative fluctuations $\chi_{_G}$ of the shear modulus can change by nearly a factor of 4 (see Fig.~\ref{fig:chi}), and that the linear size $\xi_g$ of quasilocalized excitations can change by up to a factor of 2 (see Fig.~\ref{fig:all_core_length_observables}) by varying glass stickiness alone. The relative magnitude of these effects is similar to that seen to be induced by thermal annealing~\cite{pinching_pnas}. 

\item The geometry of soft quasilocalized modes (QLMs) --- which have been recently shown~\cite{modes_prl_2020} to exist in any structural glass quenched from a melt --- can be sensitive to details of the interaction potential. In particular, we show in Sec.~\ref{sect:elastic_moduli_fluctuation} that the dilatational component of the QLMs' eigenstrains can grow from essentially zero (in non-sticky glasses) to a few tens of percents of their dominant shear eigenstrain (in sticky glasses). 

\item The physical mechanism responsible for the emergence of soft QLMs is non-universal (see Fig.~\ref{fig:freq_frac_distributions} and associated discussion). In particular, the dominant contribution to decreasing QLMs' energies in sticky glasses is shown to be associated with negative curvatures -- that are generically featured by attractive pairwise interactions. At the same time, we highlight in Sec.~\ref{sec:discussion_freq_frac} the role of \emph{large} negative curvatures of the pairwise potential in \emph{suppressing} the number fraction and \emph{increasing} the characteristic frequencies of QLMs.

\item We find a quasi-universal relation (see Fig.~\ref{fig:ag_vs_chig}) between two dimensionless quantifiers of mechanical disorder, namely the (dimensionless) prefactor $A_g$ of the nonphononic vDOS, and the quantifier $\chi_{_G}$ of elastic moduli fluctuations. This relation is shown to hold across 8 different glass models that span very large ranges of these two quantifiers.

\end{enumerate}

\section{Outlook}
\label{sect:outlook}

Our results in this work constitute a starting point for several further investigations. We first reiterate that in an accompanying paper \cite{inannealability} we show that glass stickiness affects the way thermal annealing --- in the form of deep supercooling of parent equilibrium states --- induces changes in elastic properties of simple computer glasses. 

Next, we propose that the real-space counterpart of the analysis presented in Fig.~\ref{fig:freq_frac_distributions}, that shows how low-frequency QLMs' energies are composed of stabilizing and destabilizing terms, and the latters' relative contributions, should be systematically carried out. The aim would be to reach an understanding regarding the key destabilizing mechanisms and their variation as a function of the interparticle potential properties, and as a function of the emergent populations of interactions featuring positive and negative stiffnesses and forces. 

The mechanical properties of the CSS model have been investigated in Refs.~\cite{itamar_brittle_to_ductile_pre_2011,smarajit_ductile_brittle_soft_matter_2016}, where it was shown that increasing glass stickiness leads to brittle-like failure under uniaxial tension. It would be interesting to investigate how the deformation-coupling properties of the STZs (which are a subset of the QLMs) in the CSS model, as shown in Fig.~\ref{fig:dilation_shear} and discussed in Sec.~\ref{sect:elastic_moduli_fluctuation}, affect different failure modes such as uniaxial compression or hydrostatic tension. Furthermore, it is important to establish to what extent the analogy between thermal-annealing-induced stability and glass-stickiness-induced stability generally persists in dynamic mechanical tests of sticky-sphere glasses. 

Finally, one of our key results, shown in Fig.~\ref{fig:xig_vs_chig}, strongly suggests a close connection between the (sample-to-sample) fluctuations of macroscopic shear moduli, as captured by $\chi_{_G}$, and the crossover length $\xi_g$, which also represents the core-size of STZs~\cite{pinching_pnas}. This connection should be more firmly established, with the aim of building a unifying framework that will allow to effectively quantify the degree of mechanical disorder that a given glass possesses, and compare it --- on the same footing --- with other classes of disordered materials.


\acknowledgements

We warmly thank Srikanth Sastry, Geert Kapteijns, David Richard, Eran Bouchbinder, and J. Chattoraj for fruitful discussions. We are indebted to David Richard for providing us Stillinger-Weber computer glasses and elasticity data. E.~L.~acknowledges support from the Netherlands Organisation for Scientific Research (NWO) (Vidi grant no.~680-47-554/3259). K.~G.~L gratefully acknowledges the computer resources provided by the Laboratorio Nacional de Superc\'omputo del Sureste de M\'exico, CONACYT member of the national laboratories network. M.~P.~C.~acknowledges support from the Singapore Ministry of Education throughthe Academic Research Fund (MOE2017-T2-1-066(S)). Parts of this work were carried out on the Dutch national e-infrastructure with the support of SURF Cooperative.

\appendix
\section{Definitions of observables}\label{appx:defs}

In this section we list and provide precise definitions of the physical observables we focused on in this study, and some of the methods of their measurement. We divide the observables to macroscopic and microscopic ones, in the next Subsections.

\subsection{Macroscopic elasticity}
\label{appx:macro}

\subsubsection{Elastic moduli}

We start with athermal ($T\!=\!0$) elastic moduli \cite{lutsko}; the shear modulus $G$ is defined as
\begin{equation}\label{eq-G}
    G \equiv \frac{1}{V}\frac{d^2U}{d\gamma^2}= \frac{\frac{\partial^{2}U}{\partial \gamma^{2}}-\frac{\partial^{2}U}{\partial \gamma\partial \xv} \cdot \calBold{M}^{-1}\cdot \frac{\partial ^{2}U}{\partial\xv\partial \gamma}}
    {V}\,,
\end{equation}
where $d/d\gamma$ denotes the total derivative under the constraints of mechanical equilibrium \cite{lutsko}, $\xv$ denotes particles' coordinates, $\calBold{ M}\!\equiv\!\frac{\partial^2U}{\partial\xv\partial\xv}$ is the Hessian matrix of the potential $U$, and $\gamma$ is a shear-strain parameter that parameterizes the imposed affine simple shear (in the $x$-$y$ plane) transformation of coordinates $\xv\!\to \mathBold{H}(\gamma)\cdot\xv$ with
\begin{equation}\label{shear_transformation_matrix}
\mathBold{H}(\gamma) =  \left( \begin{array}{ccc}1&\gamma&0\\0&1&0\\
0&0&1\end{array}\right)\,.
\end{equation}

The bulk modulus $K$ is defined as
\begin{equation}\label{eq-K}
    K \equiv -\frac{1}{\dbar}\frac{dp}{d\eta} = \frac{\frac{\partial^{2}U}{\partial \eta^{2}} -\dbar \frac{\partial U}{\partial\eta}- \frac{\partial^{2}U}{\partial \eta \partial \xv} \cdot \calBold{M}^{-1} \cdot \frac{\partial^{2}U}{\partial\xv\partial\eta}}{V\dbar^2}\,,
\end{equation}
where 
\begin{equation}
p\equiv-\frac{1}{V\dbar}\frac{\partial U}{\partial \eta}
\end{equation}
is the pressure, $\dbar$ is the dimension of space, and $\eta$ is an expansive-strain parameter that parameterizes the imposed affine expansive transformation of coordinates $\xv\!\to \mathBold{H}(\eta)\cdot\xv$ as
\begin{equation}\label{dilation_transformation_matrix}
\mathBold{H}(\eta) =  \left( \begin{array}{ccc}e^\eta&0&0\\0&e^\eta&0\\0&0&e^\eta\end{array}\right)\,.
\end{equation}
With the definitions of the shear and bulk moduli in hand, the Poisson's ratio $\nu$ of a 3D solid is given by
\begin{equation}\label{eq-poisson}   
    \nu \equiv \frac{3K-2G}{6K+2G} = \frac{3-2G/K}{6+2G/K}\,.
\end{equation}

\subsubsection{Characteristic pressure scale}
\label{app:p_0}

\begin{figure}[h!]
\centering
  \includegraphics[width=1.0\linewidth]{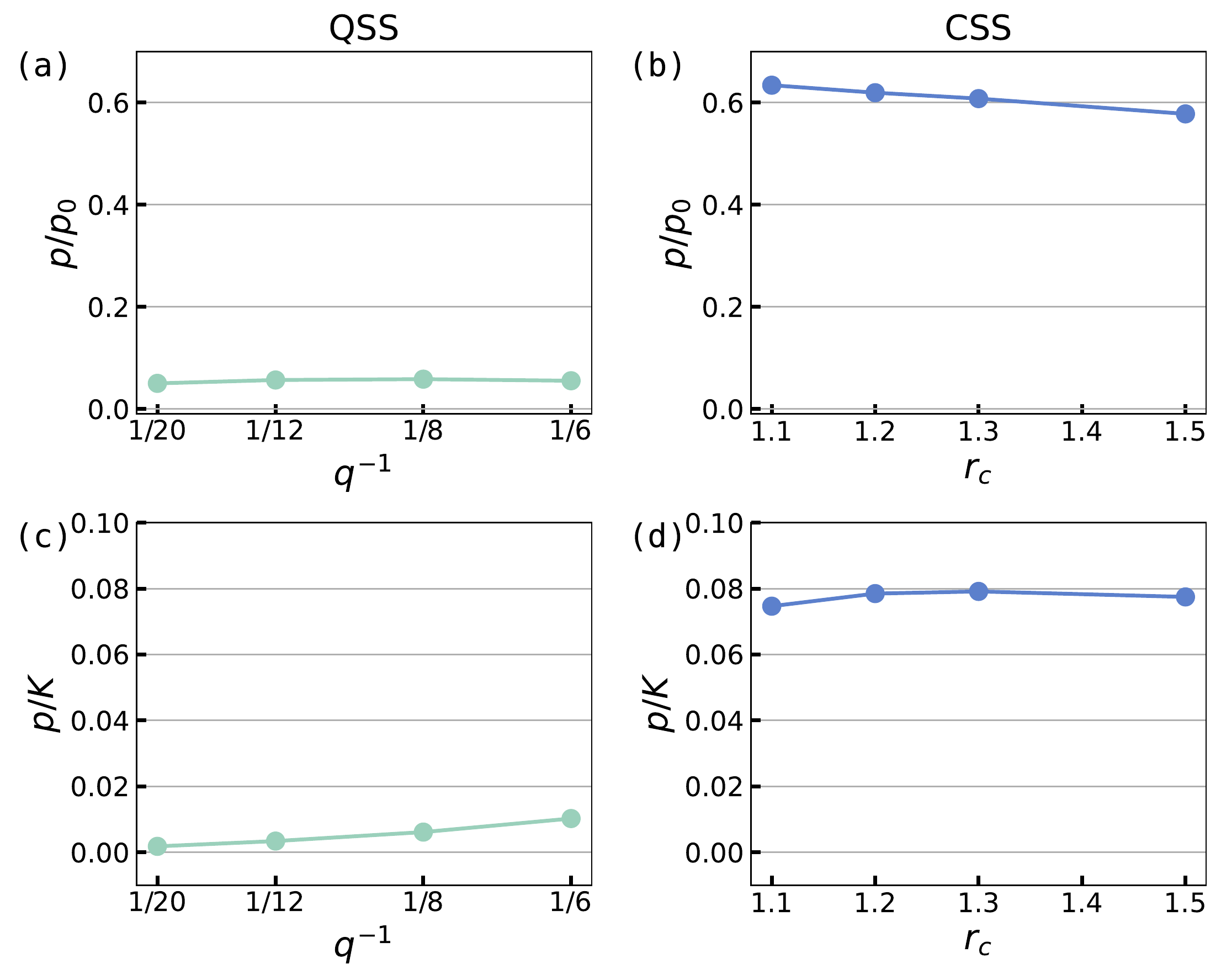}
\caption{\footnotesize Panels (a) and (b) show the dimensionless pressure $p/p_0$ for the QSS and CSS ensembles, respectively. We have tuned the density carefully in the QSS system such that $p/p_0\!\approx\!0.05$, in order to achieve maximum glass stability while maintaining a positive pressure. In the CSS system we kept the density fixed. Panels (c) and (d) show the pressure made dimensionless via rescaling by the bulk modulus $K$, for comparison.
\label{fig:pressures}}
\end{figure}

For models featuring radially-symmetric pairwise potentials --- as employed throughout this work --- the pressure can be decomposed into positive and negative contributions, as
\begin{equation}
    p = \frac{1}{V\dbar} \sum_{f_{ij}>0} f_{ij}r_{ij} - \frac{1}{V\dbar} \sum_{f_{ij}<0}(-f_{ij})r_{ij} \equiv p_+ - p_-\,.
\end{equation}
It is natural to use this decomposition to define a characteristic scale $p_0\!\equiv\!p_+ + p_-$ with respect to which the pressure can be assessed. In Fig.~\ref{fig:pressures} we report $p/p_0$ for two of the three glass ensembles investigated, in addition of the ratios $p/K$ for comparison. We assert that since characteristic forces and characteristic stiffnesses do not vary together in the different $q$ glasses, the correct way to compare these systems on the same footing is by (approximately) fixing the scale $p_0$, as we did for the QSS glasses. 

\begin{figure*}
\centering
  \includegraphics[width=0.78\linewidth]{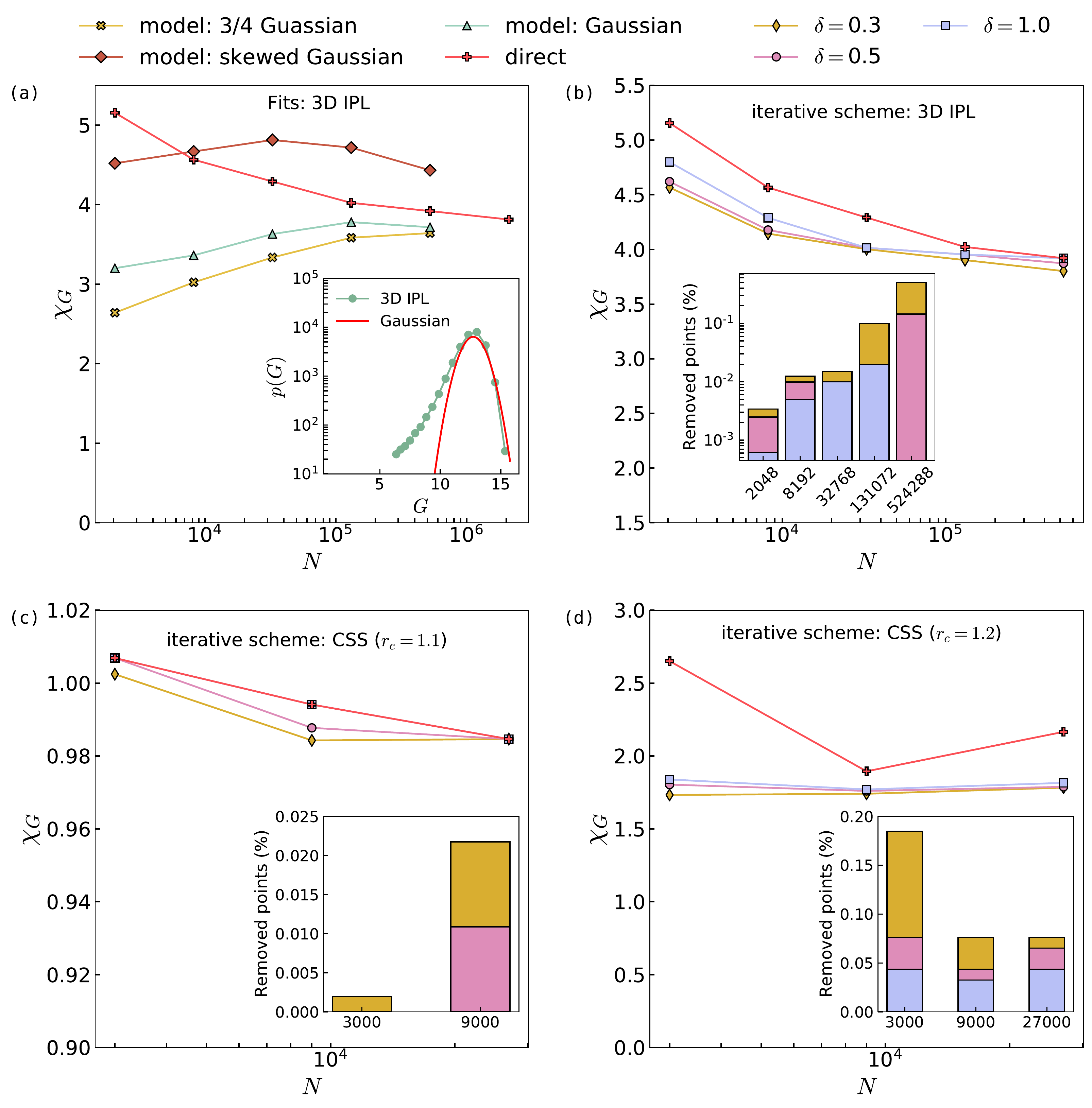}
\caption{\footnotesize Robustness of the scheme employed for estimating the disorder quantifier $\chi_{_G}$. The scheme is tested first on data  obtained for a computer glass former of soft repulsive spheres, and various glass sizes $N$; model details, glass preparation protocol and glass-ensemble sizes can be found in Ref.~\cite{finite_size_effect_modes_2020}. Panel (a) compares between the value of $\chi_{_G}$ obtained by a direct calculation, to that obtained by fitting the probability distribution function $p(G)$ to a Gaussian, a `3/4 Gaussian', and a skewed Gaussian. The inset of panel (a) demonstrates that the generic form of $p(G)$ features a non-Gaussian tail towards negative $G$'s~\cite{scattering_arXiv_2020}. Panels (b), (c), and (d) test the $\delta$-dependence of our scheme applied to (b) the same system shown in (a), to the CSS model with $r_c=1.1$ (c), to the CSS model with $r_c=1.2$ (d). The insets of panels (b),(c), and (d) report the percentage of outliers removed from the original data set by our scheme, for each $\delta$ (color coded as the legend). We conclude that the most effect percentage $\delta\!=\!1\%$. 
\label{fig:outlier_rmval}}
\end{figure*}

\subsubsection{Sample-to-sample elastic moduli fluctuations}
\label{app:chi}

We also consider two dimensionless, $N$-independent measures of the sample-to-sample fluctuations of shear and bulk elastic moduli, defined respectively as 
\begin{equation}\label{eq:appendix_chi_def}
    \chi_{_G} = \frac{\sqrt{N\langle (G - \langle G \rangle)^2 \rangle}}{\langle G \rangle} \ \  \mbox{and} \ \  \chi_{_K} = \frac{\sqrt{N\langle (K - \langle K \rangle)^2 \rangle}}{\langle K \rangle}\,,
\end{equation}
where $\langle\bullet\rangle$ denotes an ensemble average (these definitions are also spelled out in Eq.~(\ref{eq:chi}) of Sect.~\ref{sect:macro}). Estimating the sample-to-sample variance of $G$ (appearing in the definition of $\chi_{_G}$ above) is difficult, since its probability distribution is known to feature strong finite size effects~\cite{scattering_arXiv_2020}, similar to those discussed at length in Refs.~\cite{scattering_arXiv_2020,finite_size_effect_modes_2020,corrado_dipole_statistics_2020}. In particular, it is common to observe outliers with large and negative $G$ in small glasses, that can corrupt the estimated variance in our finite data set. 

For these reasons we opt for estimating $\chi_{_G}$ using the following approach; we start by calculating $\chi_{_G}$ directly using the entire raw data set, following the definition as spelled out in Eq.~\ref{eq:appendix_chi_def}. Then:
\begin{enumerate}
    \item For each data point $G_i$ we calculate $\chi_{_G}^{i}$, which is the same as $\chi_{_G}$ but computed while \emph{excluding} the $i$'th data point $G_i$ from the calculation. Once $\chi_{_G}^{i}$ is obtained, the $i$'th data point $G_i$ is returned to the data set.
    \item We define a percent difference $\delta_i\!\equiv\!100\!\times\!(\chi_{_G}\! -\!\chi_{_G}^{i})/\chi_{_G}$ for each data point $i$.
    \item The data point with the \emph{largest} percent difference $\delta_i$ is \emph{permanently} removed from the total data set, and $\chi_{_G}$ is re-calculated.
    \item Goto 1.
\end{enumerate}
The scheme described above is repeated, until the maximal percent difference falls below a fixed percentage $\delta$, which we set at 1\%, as explained and motivated in Fig.~\ref{fig:outlier_rmval} and its caption.

To test the scheme proposed above for estimating $\chi_{_G}$, we compare the result of our estimations with other approaches: a full Gaussian fit, a `3/4 Gaussian' fit, and a skewed Gaussian \cite{azzalini2013skew}, all using the built-in model from the LMFIT python library~\cite{LMFIT_python}. In addition, we test the effect of the threshold percentage $\delta$ employed. The results of these tests are displayed in Fig.~\ref{fig:outlier_rmval} and explained it its caption. Our proposed scheme for handling the finite-size-induced noise in $\chi_{_G}$, which is similar in spirit to the Jack-Knife method, was also employed for estimating~$\chi_{_K}$.

\subsection{Microscopic elasticity}
\label{app:micro}

\subsubsection{Vibrational density of states}

The vibrational density of states (vDOS) is defined as
\begin{equation}
    D(\omega) = \frac{1}{N}\bigg<\sum_\ell\delta(\omega - \omega_\ell)\bigg>\,,
\end{equation}
where $\omega_\ell$ is the vibrational frequency associated with the vibrational mode $\psiv^{(\ell)}$ that solves the eigenvalue equation
\begin{equation}
\calBold{M}\cdot\psiv^{(\ell)} = \omega_\ell^2 \psiv^{(\ell)}\,,
\end{equation}
assuming all masses are identical and equal to unity. It is known that $D(\omega)\!=\!A_g\omega^4$ as $\omega\!\to\!0$ in structural glasses quenched from a melt \cite{modes_prl_2016,modes_prl_2018,modes_prl_2020}, and see also Fig.~\ref{fig:prefactor} of the main text. In the same figure, the prefactor $A_g$, which has units of a frequency$^{-5}$, is measured and reported for our glass ensembles.

\subsubsection{Mesoelastic length and frequency scales}
\label{xi_g_and_omega_g_appendix}

In order to extract the lengthscale that characterizes soft, quasilocalized modes, we follow the procedure introduced in \cite{pinching_pnas}; first, we impose a force dipole to a pair of interacting particles, of the form
\begin{equation}
\dv_{ij} = \frac{\partial r_{ij}}{\partial\xv}\,.
\end{equation}
The normalized response to this force is 
\begin{equation}
\zv_{ij} = \frac{\calBold{M}^{-1}\cdot\dv_{ij}}{\sqrt{\dv_{ij}\cdot\calBold{M}^{-2}\cdot\dv_{ij}}}\,.
\end{equation}
We define a correlation function $c(r)$ as
\begin{equation}\label{decay_function_definition}
    c(r) \equiv \bigg< \mbox{median}_{r_{ij,k\ell}\approx r}(\zv_{ij}\cdot \dv_{k\ell})^2 \bigg>\,,
\end{equation}
where the median is taken over all pairs $k,\ell$ whose distance to the $i,j$ pair is $r$, and the average is taken over independent samples. Examples of the functions $c(r)$ are shown in Fig.~\ref{fig:all_core_length_observables}a-c of the main text.

\begin{figure}[h!]
\centering
  \includegraphics[width=0.9\linewidth]{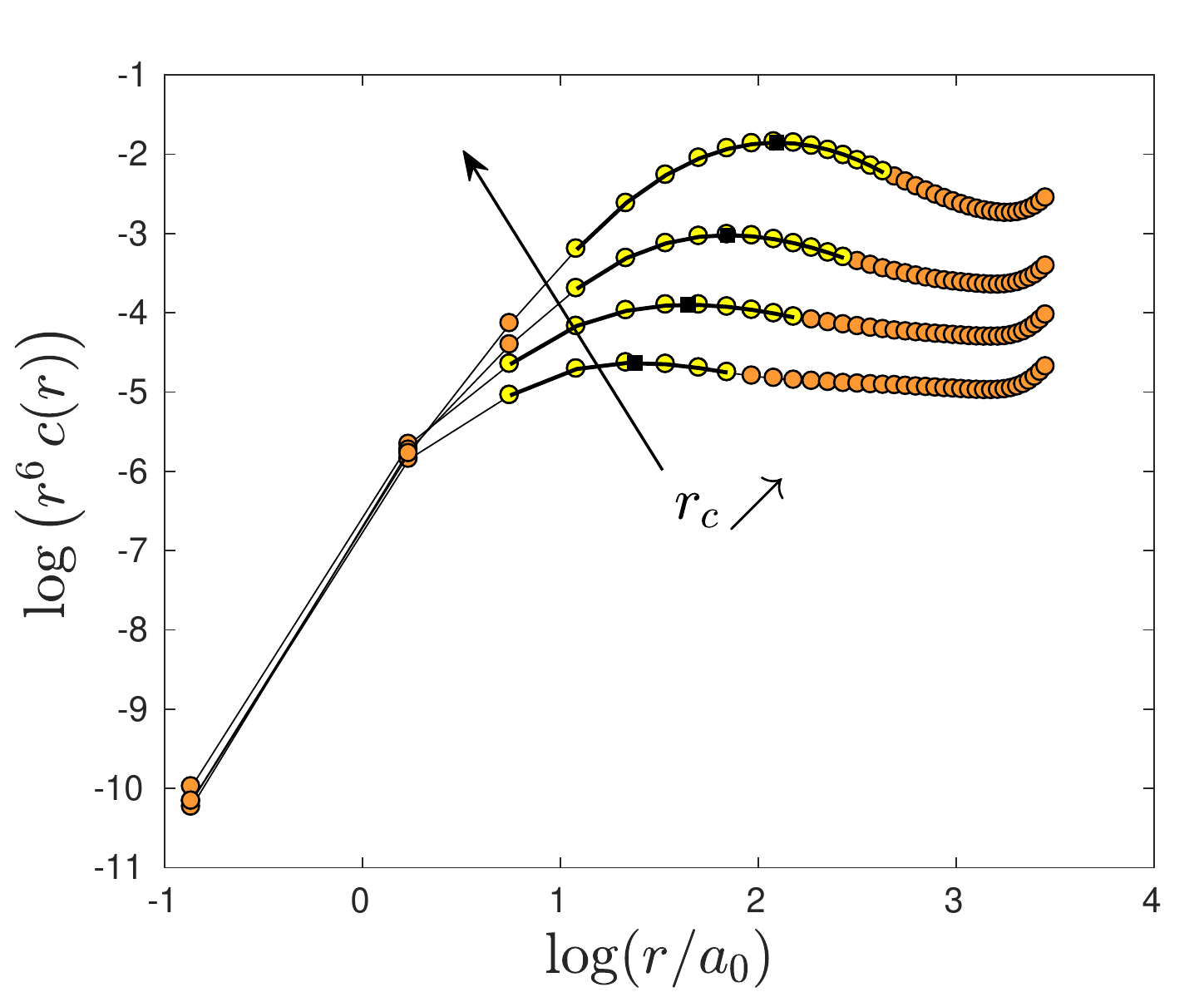}
\caption{Decay functions $c(r)$ (defined in Eq.~(\ref{decay_function_definition})) measured in the CSS system with different cutoffs $r_c$, factored by $r^{6}$ and plotted \emph{vs.}~$r/a_{0}$, where $r$ is the distance from the applied local force dipole, and $a_0$ is a characteristic interparticle distance. The big arrow indicates the direction of increasing interaction cutoff $r_c$. With this figure we illustrate how the length at which the crossover to the continuum scaling, $\xi_{g}$, is extracted. We fit a 3rd degree polynomial to the (logarithm base 10 of the) signal in the bump range (highlighted with yellow markers). The fit is represented by the black curves on top of the yellow markers, and the full black squares mark the extracted values for $\xi_{g}$.
\label{fig:length_extraction}}
\end{figure}

Continuum elasticity tells us that, at large $r$, $c(r)\!\sim\!r^{-2\dbar}$ \cite{breakdown}; we extract a mesoscopic length $\xi_g$ by finding the maximum of the product $r^6c(r)$, as shown in Fig.~\ref{fig:length_extraction}. With a mesoscopic length in hand, we follow \cite{pinching_pnas} and define a characteristic frequency scale of quasilocalized modes as $\omega_g\!\equiv\!2\pi c_s/\xi_g$. Finally, using the prefactor~$A_g$ of the vDOS, and the frequency scale $\omega_g$, the number density of quasilocalized modes is obtained via ${\cal N}\!\equiv\!A_g\omega_g^5$, as discussed in \cite{pinching_pnas}. $\omega_g$ and ${\cal N}$ measured in our model glasses are shown in Fig.~\ref{fig:wg_N_QLE} in the main text. 

\subsubsection{Deformation coupling of quasilocalized modes}
\label{app:eigenstrains}

Studying QLMs' properties in computer glasses via a harmonic analysis can be challenging, due to the narrow set of conditions in which such an analysis is able to reveal those properties \cite{phonon_widths,pinching_pnas}. In particular, in situations in which $A_g$ is small (e.g.~in the CSS systems with small $r_c$, see Fig.~\ref{fig:prefactor}), revealing QLMs by harmonic analyses is difficult, and requires unreasonably large ensemble sizes. For this reason, the deformation patterns associated with QLMs are most conveniently studied using the nonlinear quasilocalized excitations framework put forward in \cite{SciPost2016, episode_1_2020}. In this framework, nonlinear QLMs $\piv$ solve the algebraic equation
\begin{equation}
    \calBold{M}\cdot\piv = \frac{\calBold{M}:\piv\piv}{\mathBold{U''''}::\piv\piv\piv\piv}\,\mathBold{U''''}:\!\cdot\,\piv\piv\piv\,,
\end{equation}
where $\mathBold{U''''}\!\equiv\!\frac{\partial^4U}{\partial\xv\partial\xv\partial\xv\partial\xv}$ is the rank-4 tensor of derivatives of the potential energy with respect to coordinates, and $::,:\!\cdot$ denote quadruple and triple contractions (over $N\!\times\!\dbar$ components), respectively.
This framework currently allows to compute the one of the softest quasilocalized modes given a computer glass, regardless of how its frequency relates to phonon frequencies, and of the degree of hybridization of \emph{harmonic}, quasilocalized modes with phonons. We followed the protocol described in \cite{episode_1_2020} to calculate a single soft nonlinear QLM in each glassy sample. 

Recently, a set of tools to extract the core properties of QLMs was developed \cite{avraham_arXiv}. For a given QLM $\piv$, its coupling to dilatational and shear deformations can be extracted as follows \cite{avraham_arXiv}; we define the tensor $\calBold{F}_{\piv}$ as
\begin{equation}
    \calBold{F}_{\piv} = \frac{\partial^2U}{\partial\mathBold{\epsilon}\partial\xv}\cdot\piv\,,
\end{equation}
where $\mathBold{\epsilon}$ is the strain tensor that quantifies the geometry and amplitude of imposed deformations. $\calBold{F}_{\piv}$ is decomposed into a deviatoric (traceless) and isotropic terms as $\calBold{F}_{\piv}\!=\!\calBold{F}_{\piv}^{\mbox{\footnotesize dev}}\!+\!\calBold{F}_{\piv}^{\mbox{\footnotesize iso}}$, where $\calBold{F}_{\piv}^{\mbox{\footnotesize iso}}\!=\!\calBold{I}\mbox{Tr}\big(\calBold{F}_{\piv}\big)/\dbar$ ($\calBold{I}$ is the identity tensor) and $\calBold{F}_{\piv}^{\mbox{\footnotesize dev}}\!=\!\calBold{F}_{\piv}-\calBold{F}_{\piv}^{\mbox{\footnotesize iso}}$. Without loss of generality, if the eigenvalue of $\calBold{F}_{\piv}^{\mbox{\footnotesize dev}}$ with the largest absolute magnitude is \emph{negative}, we switch $\piv\!\to\!-\piv$ (note that $\piv$ is defined up to a sign~\cite{SciPost2016,episode_1_2020}). Denoting then the largest eigenvalue of $\calBold{F}_{\piv}^{\mbox{\footnotesize dev}}$ by $\lambda_{\mbox{\scriptsize max}}$, the ratio of dilation to shear coupling is given by \cite{avraham_arXiv}
\begin{equation}\label{eq:dil_shear_eigenstrains}
\frac{\mbox{dilation}}{\mbox{shear}} \equiv \frac{\mbox{Tr}\big(\calBold{F}_{\piv}\big)/\dbar}{\lambda_{\mbox{\scriptsize max}}}\,.
\end{equation}
This ratio is reported in Sect.~\ref{sect:elastic_moduli_fluctuation} for ensembles of nonlinear modes calculated in our glasses (one mode per glass).




\section{Interaction potentials' coefficients}
\label{app:coefficients}

We provide a mathematical expression to obtain the coefficients $c_{\mbox{\tiny $2\ell$}}$ as a function of the exponent q in the $\varphi_{\mbox{\tiny QSS}}$ potential (see Eq.~(\ref{eq:qstickypot}). The expression reads as:

\begin{equation}\label{eq:qexponents}
    \begin{pmatrix}
c_{\mbox{\tiny $0$}}(q) \\
c_{\mbox{\tiny $2$}}(q) \\
c_{\mbox{\tiny $4$}}(q) 
\end{pmatrix}
\!=\!
    \begin{pmatrix}
 \!\frac{1}{4} x_{c}^{-2q}(2+q)(4 q x_{c}^{q}-2(q+1)) \!\\
\!-\frac{1}{2} q x_{c}^{-2(q+1)}((4+q)x_{c}^{q}-2(q+2)) \!\\
\!\frac{1}{4} q x_{c}^{-2(2+q)}((2+q)x_{c}^{q}-2(q+1))\!
\end{pmatrix}
\end{equation}

Additionally, table \ref{tab:CSSoeff}, provides the coefficients that ensure the first and second derivatives of the sticky spheres interaction potential, $\varphi_{\mbox{\tiny SS}}$ to be smooth at the dimensionless interaction cutoff $x_{c}\!=\!r_{c}\!\times\! x_{\mbox{\tiny min}}$. 

\begin{table}
\caption{\label{tab:CSSoeff}
Sticky Spheres potential coefficients.}
\begin{ruledtabular}
\begin{tabular}{ccc}
$r_{c} = 1.1$\\
\hline
a & $-2065.398715559462$\\
b & $-6775.790387829389$ \\
$c_{\mbox{\tiny $0$}}$ & $-25758.833604839492$ \\
$c_{\mbox{\tiny $2$}}$ & $35758.32100717132$ \\
$c_{\mbox{\tiny $4$}}$ & $-17961.4634217834$\\
$c_{\mbox{\tiny $6$}}$ & $3172.8363373043844$\\
\hline
$r_{c} = 1.15$\\
\hline
a & $-360.56276228220503$\\
b & $-1087.3468598689665$ \\
$c_{\mbox{\tiny $0$}}$ & $-3686.241663716159$ \\
$c_{\mbox{\tiny $2$}}$ & $4884.979422475693$ \\
$c_{\mbox{\tiny $4$}}$ & $-2333.6422601463023$ \\
$c_{\mbox{\tiny $6$}}$ &$390.72335017374587$\\
\hline
$r_{c} = 1.2$\\
\hline
a & $-106.991613526652$ \\
b & $-304.918469059567$ \\
$c_{\mbox{\tiny $0$}}$& $-939.388037994211$ \\
$c_{\mbox{\tiny $2$}}$& $1190.70962256002$ \\
$c_{\mbox{\tiny $4$}}$& $-541.3001315875512$ \\
$c_{\mbox{\tiny $6$}}$& $85.86849369147127$\\
\hline
$r_{c} = 1.3$\\
\hline
a & $-17.7556513878655$ \\
b & $-50.37332289908061$ \\
$c_{\mbox{\tiny $0$}}$& $-138.58271673010657$ \\
$c_{\mbox{\tiny $2$}}$& $161.71576064627635$ \\
$c_{\mbox{\tiny $4$}}$& $-66.7252832098764$ \\
$c_{\mbox{\tiny $6$}}$& $9.50283097488097$\\
\hline
$r_{c} = 1.5$\\
\hline
a & $1.1582440286928275$ \\
b & $-2.2619482444770567$ \\
$c_{\mbox{\tiny $0$}}$ & $-12.414700446492716$ \\
$c_{\mbox{\tiny $2$}}$ & $12.584354590303674$ \\
$c_{\mbox{\tiny $4$}}$ & $-4.320508006050397$ \\
$c_{\mbox{\tiny $6$}}$ & $0.49862551162881885$\\
\end{tabular}
\end{ruledtabular}
\end{table}



\clearpage

\begin{thebibliography}{122}
\bibitem [{foo({\natexlab{b}})}]{footnote_ncal}%
  \BibitemOpen
  \href@noop {} {\bibfield  {journal} {\bibinfo  {journal} {The precise value
  of the quasi-universal $\cal{N}$ depends on how QLMs' characteristic
  frequency $\omega_g$ is obtained. In~\cite{inannealability} it is shown that
  $\omega_g$ estimated via the response to local force dipole (as explained in,
  e.g., \cite{cge_paper}) agrees well --- up to an overall numerical factor of
  order unity --- with the $\xi_g$-based estimation as done here}\ } }\BibitemShut {NoStop}%
\bibitem [{foo({\natexlab{c}})}]{footnote2}%
  \BibitemOpen
  \href@noop {} {\bibfield  {journal} {\bibinfo  {journal} {In cases where the
  interaction-network topology is fixed, e.g.~networks of Hookean springs as
  considered in \cite{breakdown}, the destabilizing contribution $k_-$ due to
  negative stiffnesses might be absent}\ }}\BibitemShut {NoStop}%
\end{thebibliography}
%

\end{document}